\documentclass[namedreferences]{solarphysics}
\usepackage[optionalrh, natbib]{spr-sola-addons}
\usepackage{graphicx}
\usepackage{url}
\usepackage[breaklinks=true]{hyperref}

\newcommand{\aap}{    {\it Astron. Astrophys.}}
\newcommand{\apj}{    {\it Astrophys. J.}}
\newcommand{\apss}{   {\it Astrophys. Space Sci.}} 
\newcommand{\araa}{		{\it Annu. Rev. Astron. Astrophys.}}
\newcommand{\pasj}{   {\it Publ. Astron. Soc. Japan}}
\newcommand{\solphys}{{\it Solar Phys.}}
\newcommand{\sovast}{ {\it Soviet  Astron.}} 
\newcommand{\ssr}{    {\it Space Sci. Rev.}} 

\begin{document}
\begin{article}
\begin{opening}
\title{Simulations of Gyrosynchrotron Microwave Emission from an Oscillating 3D Magnetic Loop}
\author{A.A.~\surname{Kuznetsov}$^{1}$\sep
        T.~\surname{Van Doorsselaere}$^{2}$\sep
				V.E.~\surname{Reznikova}$^{2, 3}$}
\runningtitle{Microwave emission from an oscillating 3D magnetic loop}
\runningauthor{Kuznetsov, Van Doorsselaere, and Reznikova}        
\institute{$^{1}$ Institute of Solar-Terrestrial Physics, Irkutsk 664033, Russia
                  e-mail: \url{a_kuzn@iszf.irk.ru}\\
           $^{2}$ CmPA, Department of Mathematics, KU Leuven, BE-3001 Leuven, Belgium
					        e-mail: \url{tom.vandoorsselaere@wis.kuleuven.be}\\
					 $^{3}$ Radiophysical Research Institute (NIRFI), Nizhny Novgorod 603950, Russia}
\begin{abstract}
Radio observations of solar flares often reveal various periodic or quasi-periodic oscillations. Most likely, these oscillations are caused by magnetohydrodynamic (MHD) oscillations of flaring loops which modulate the emission. Interpretation of the observations requires comparing them with simulations.
We simulate the gyrosynchrotron radio emission from a semi-circular (toroidal-shaped) magnetic loop containing sausage-mode MHD oscillations. The aim is to detect the observable signatures specific to the considered MHD mode and to study their dependence on the various source parameters.
The MHD waves are simulated using a linear three-dimensional model of a magnetized plasma cylinder; both standing and propagating waves are considered. The curved loop is formed by replicating the MHD solutions along the plasma cylinder and bending the cylinder; this model allows us to study the effect of varying the viewing angle along the loop. The radio emission is simulated using a three-dimensional model and its spatial and temporal variations are analyzed. We consider several loop orientations and different parameters of the magnetic field, plasma, and energetic electrons in the loop.
In the model with low plasma density, the intensity oscillations at all frequencies are synchronous (with the exception of a narrow spectral region below the spectral peak). In the model with high plasma density, the emission at low frequencies (where the Razin effect is important) oscillates in anti-phase with the emissions at higher frequencies. The oscillations at high and low frequencies are more pronounced in different parts of the loop (depending on the loop orientation). The layers where the line-of-sight component of the magnetic field changes sign can produce additional peculiarities in the oscillation patterns.
\end{abstract} 
\end{opening}

\section{Introduction}
Waves and oscillations are a common phenomeno in solar flares. They are believed to reveal themselves as quasi-periodic pulsations in spatially unresolved observations at different wavelengths \citep[see, \textit{e.g.}, the review of][]{nak09} and have been directly detected using the EUV observations with high spatial resolution \citep{nak99, asc99, dem12}. In particular, the observed periodicities of the quasi-periodic pulsations in the radio emission, hard X-rays, and $\gamma$-rays (from a fraction of a second to several minutes) agree with the expected magnetohydrodynamic (MHD) timescales which strongly suggests a close relationship between the pulsations and the MHD waves and oscillations \citep{nak09, dem12}. It should be noted, however, that the solar quasi-periodic pulsations are usually superimposed on a rapidly varying background, so that recovering their parameters is not easy; some previously reported conclusions may be unreliable due to the methodology used \citep{gru11}.

Observations of the waves and oscillations in the flares can be used to diagnose the quasi-periodic magnetic reconnection processes that are believed to be the most probable source of these oscillations \citep[\textit{e.g.},][]{kli00}. The MHD waves and oscillations can also be used to diagnose the coronal plasma and magnetic field parameters \citep[``coronal seismology''; see, \textit{e.g.},][]{nak05}. Moreover, the waves and oscillations are believed to play a key role in the energy transfer both in the quiet corona \citep[\textit{e.g.},][]{erd06} and in the flares \citep{fle08}; they can modulate and/or trigger the magnetic reconnection processes \citep{che06, doy06, nak06} and thus affect greatly the behaviour of unstable active regions and development of the solar flares. 

One of the potential observable manifestations of MHD waves is quasi-periodic pulsations of radio emission \citep[\textit{e.g.},][]{asc87, nak09, kup10}, because these waves can modulate both the coherent and incoherent radio emission mechanisms. Interpretation of the observed quasi-periodic pulsations requires comparing the observations with simulations in order to determine the origin of the pulsations (because they can be caused not only by the MHD waves but also, \textit{e.g.}, by quasi-periodic injection of energetic particles); if the oscillations are due to waves, it allows us to identify the wave mode and other characteristics of the MHD waves. In turn, identification of the wave mode is of high importance for the development of the coronal seismology. As a rule, the basic characteristics of the pulsations (such as the period) have no unique interpretation and therefore some more subtle features (such as the relations between the pulsations observed in different spectral channels) have to be analyzed \citep{gfle08, mos12}.

Perhaps the best tool to study the physical parameters and processes in solar flares is their microwave emission, because it is produced mainly due to the incoherent gyrosynchrotron radiation mechanism which is (a) relatively simple and well studied \citep{mel68, ram69} and (b) highly sensitive to the magnetic field and plasma parameters. There are a lot of papers dedicated to simulating the gyrosynchrotron emission in solar flares. In particular, two-dimensional (2D) models \citep[\textit{e.g.},][]{ali84, kle84, pre88, kuc93, kun01} and three-di\-men\-si\-o\-nal (3D) models \citep[\textit{e.g.},][]{pre92, bas98, nin00, sim06, cos13} have been employed to account for the structure of the emitting region. Some simulations involve the magnetic field structure obtained by extrapolating the observed photospheric magnetograms \citep[\textit{e.g.},][]{gar13, kuz14}, take into account anisotropy of the emitting electrons \citep[\textit{e.g.},][]{gfle03, alt08, sim10, kuz11} and/or evolution of the electron distributions during flares \citep[\textit{e.g.},][]{tza08, rez09, kuz10, skuz12}. However, possible modulation of the emission by MHD waves has been studied only in a few works and by using relatively simple models \citep{kop02, nak06b, rez07, mos12}. Recently, \cite{rez14} have simulated the gyrosynchrotron radio emission from an oscillating magnetic cylinder (with standing sausage waves). It was found that even for the simple model considered (a straight cylinder), calculating the emission requires 3D simulations that also take into account the spatial structure of the MHD waves, because the results can differ from simple estimates for a spatially homogeneous source \citep[\textit{cf.}][]{rez07, mos12}. At the same time, another complication arises from the fact that the solar flaring loops are (in contrast to the model of \citet{rez14}) evidently curved. This work is dedicated to simulating the radio emission from such curved loops.

Simulation of MHD waves in flaring loops is a nontrivial task. Oscillation modes of a straight overdense cylinder of infinite length are well studied \citep[\textit{etc.}]{edw83, nak05}, and the numerical simulations can be greatly simplified by using the analytical results. However, these results are not applicable to curved loops \citep{vand04}. Variation of the plasma and magnetic field parameters as well as of the loop radius with height presents even more difficulties \citep{and05, and11}. Therefore in this work we use a simplified model in which an oscillating cylinder is artificially bent to form a semi-circular loop without actually using the wave relations in a semi-torus; the magnetic field, plasma density and temperature, and the loop radius (in the equilibrium state, \textit{i.e.}, without oscillations) are assumed to be constant along the loop. The main aim of this model is to study the effect of varying the viewing angle (\textit{i.e.}, the angle between the local magnetic field and the line of sight) along the loop. Although such a model cannot correctly describe the entire loop, it can reproduce the effects of the MHD oscillations in local parts of the loop, where the effects of the loop curvature and inhomogeneity on the MHD wave parameters are negligible. We consider the incoherent gyrosynchrotron radiation of energetic electrons which makes the main contribution to the flare emission in the microwave range.

The simulation model is described in Section \ref{model}. The simulation results are presented and discussed in Section \ref{results}. The conclusions are drawn in Section \ref{conclusion}.

\section{Model}\label{model}
\subsection{MHD Waves}
We use the model described earlier by \citet{zaj75}, \citet{edw83}, \citet{ant13}, \textit{etc.} In this model, there is an overdense cylinder with the magnetic field directed along its axis. The plasma density and temperature inside the cylinder are higher than outside, while the magnetic field strength inside the cylinder is lower than outside, to maintain the pressure balance. We consider the azimuthally symmetric case when all parameters depend only on the distance along the cylinder axis $z$ and the distance from this axis $r$. In an equilibrium case, the magnetic field, plasma density, and temperature are assumed to be constant inside and outside the cylinder; they experience a sharp jump at the cylinder boundary (at $r=R$).

The formulae describing the MHD disturbances are given, e.g., in the works of \citet{zaj75}, \citet{edw83}, \citet{ant13}, and \citet{rez14}; see especially the paper of \citet{rez14} where the solutions of those equations are presented for the same conditions as used in this work. Therefore we describe the properties of the MHD waves only briefly. We consider the sausage mode -- the compressible azimuthally symmetric ($m=0$) oscillations that affect both the magnetic field and plasma density and temperature. We consider both the standing and propagating waves which are described by the relations
\begin{equation}
A_{\mathrm{stand}}\propto\sin(\omega t)\sin(kz+\varphi_A),\quad
A_{\mathrm{prop}}\propto\sin(\omega t-kz+\varphi_A),
\end{equation}
respectively, where $A$ is the wave amplitude (\textit{e.g.}, in velocity), $\omega$ and $k$ are the wave frequency and wavenumber, respectively, and $\varphi_A$ is the initial phase. 

\begin{figure}
\centerline{\includegraphics{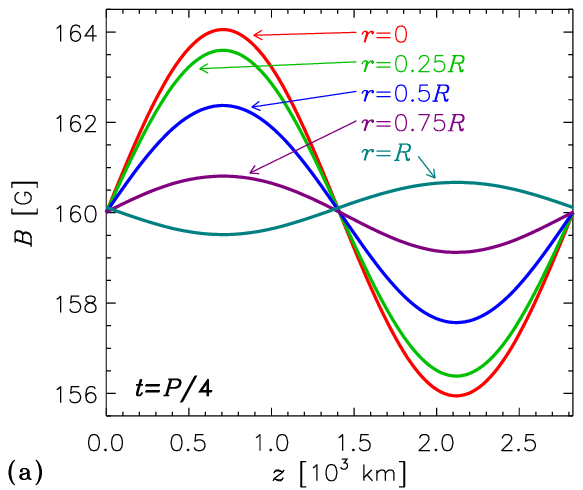}
\includegraphics{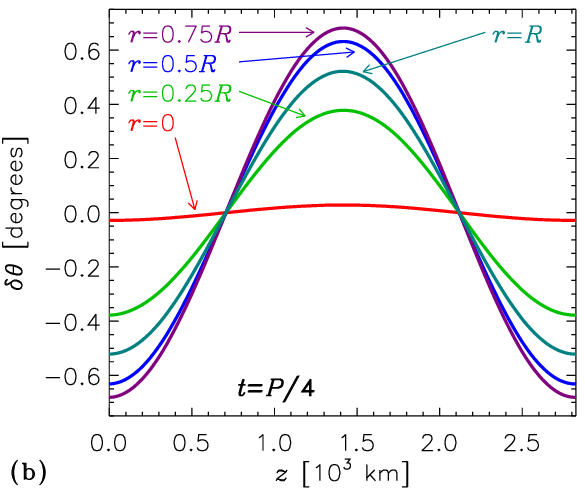}}
\caption{Profiles of the magnetic field strength (a) and direction relative to the tube axis (b) in the oscillating magnetic tube at different distances from the tube axis. The standing wave, low-density model (see Table \protect\ref{TabMHD}) is considered; one wavelength is shown. Profiles of the plasma density and temperature are similar to those of the magnetic field.}
\label{FigMHD}
\end{figure}

Figure \ref{FigMHD} presents an example of how the parameters vary with the coordinates in a standing wave (the oscillation phase is chosen to provide the maximum deviation from equilibrium). In general, the axial magnetic field $B_z$, plasma density $n_0$, and temperature $T_0$ vary in phase and their variations are maximal at the cylinder axis (we are interested only in the parameters inside the cylinder, because, as shown below, this is the region where the radio emission is assumed to be generated). Variations of the above parameters decrease with increasing $r$ and become zero at some radial distance $r_*$ nearby the cylinder boundary ($r_*\simeq 0.9R$ for the parameters considered here). Then the amplitude of the variations increases slightly, but they are now in anti-phase with the variations of the same parameters in the inner part of the cylinder. Variations of the radial magnetic field $B_r$ are shifted by $1/4$ period with respect to other parameters; they are zero at the cylinder axis and reach their maximum at $r=r_*$ (without the phase reversal at that radial distance). Since we consider low-amplitude oscillations, we have $|B_r|\ll |B_z|$ (we remind that $B_z$ includes here both the equilibrium magnetic field $B_0$ and the perturbed field component $\delta B_z$, \textit{i.e.}, $B_z=B_0+\delta B_z$). The radial component $B_r=\delta B_r$ makes almost no contribution to the resulting magnetic field strength; however, it affects slightly the magnetic field direction causing it to deviate from the cylinder axis. Variations of the cylinder radius due to the MHD oscillations are very small and do not affect the radio emission.

The propagating sausage wave looks like the structure shown in Figure \ref{FigMHD} moving along the $z$ axis. All other properties of this wave (including the phase relations between different parameters) are the same as for the standing wave.

\begin{table}
\caption{Parameters of the MHD models: a) cylinder radius ($R$), magnetic fields ($B$), plasma densities ($n_0$), and temperatures ($T_0$) inside (``in'') and outside (``out'') the cylinder in the equilibrium state, concentration of the energetic electrons ($n_{\mathrm{b}}$) inside the cylinder in the equilibrium state; b) wavelength ($\lambda$) and period ($P$) of the sausage mode, maximum deviations of the magnetic field strength ($\delta B$), magnetic field direction ($\delta\theta$), plasma density ($\delta n_0$), plasma temperature ($\delta T_0$),  and concentration of the energetic electrons ($\delta n_{\mathrm{b}}$) inside the cylinder from the respective equilibrium values.}
\label{TabMHD}
\renewcommand{\tabcolsep}{3.6em}
\begin{tabular}{lcc}
\hline
Model & Low-density & High-density\\
\hline
\multicolumn{3}{c}{\sl Equilibrium cylinder parameters}\\
$R$ [km] & 1000 & 1000\\
$B_{\mathrm{in}}$ [G] & 160 & 50\\
$B_{\mathrm{out}}$ [G] & 161 & 56\\
$n_{0\,\mathrm{in}}$ [$\textrm{cm}^{-3}$] & $4\times 10^9$ & $10^{10}$\\
$n_{0\,\mathrm{out}}$ [$\textrm{cm}^{-3}$] & $10^9$ & $3\times 10^9$\\
$T_{0\,\mathrm{in}}$ [K] & $10^7$ & $10^7$\\
$T_{0\,\mathrm{out}}$ [K] & $2\times 10^6$ & $2\times 10^6$\\
$n_{\mathrm{b}\,\mathrm{in}}$ [$\textrm{cm}^{-3}$] & $10^8$ & $2.5\times 10^8$\\
\multicolumn{3}{c}{\sl Wave parameters}\\
$\lambda$ [km] & 2825 & 2825\\
$P$ [s] & 0.35 & 1.5\\
$(\delta B_{\mathrm{in}})_{\max}$ [G] & $4.05$ & $2.78$\\
$(\delta\theta_{\mathrm{in}})_{\max}$ & $0.69^{\circ}$ & $1.52^{\circ}$\\
$(\delta n_{0\,\mathrm{in}})_{\max}$ [$\textrm{cm}^{-3}$] & $1.0\times 10^8$ & $6.0\times 10^8$\\
$(\delta T_{0\,\mathrm{in}})_{\max}$ [K] & $1.7\times 10^5$ & $4.1\times 10^5$\\
$(\delta n_{\mathrm{b}\,\mathrm{in}})_{\max}$ [$\textrm{cm}^{-3}$] & $2.5\times 10^6$ & $1.5\times 10^7$\\
\hline
\end{tabular}
\end{table}

We consider two sets of model parameters, the same as have been used in the work of \citet{rez14}; the parameters are listed in Table \ref{TabMHD}. The parameters were chosen to provide the same wavelength $\lambda$ for the given cylinder radius $R$. In the calculations below, we use mainly the low-density model because those conditions are more favorable for generation of the gyrosynchrotron radio emission. In the high-density model, the emission intensity is usually lower; at low frequencies, the emission becomes strongly affected by the Razin effect (see Section \ref{razin}).

\subsection{Source Geometry}
\begin{figure}
\centerline{\includegraphics{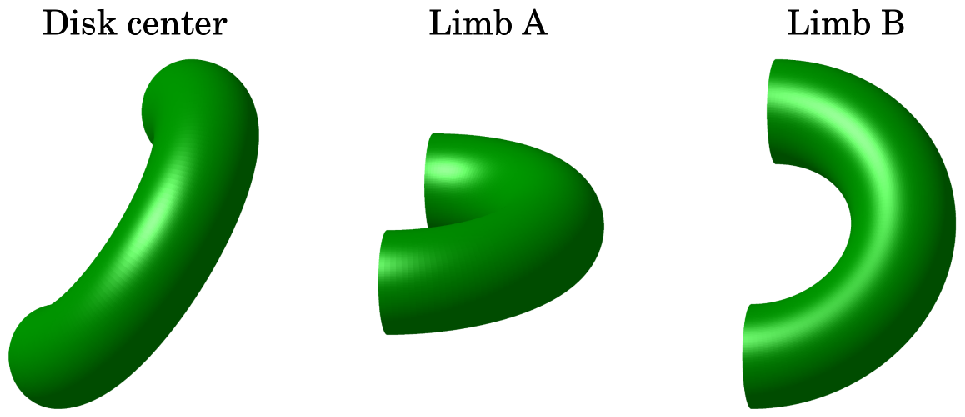}}
\centerline{\includegraphics{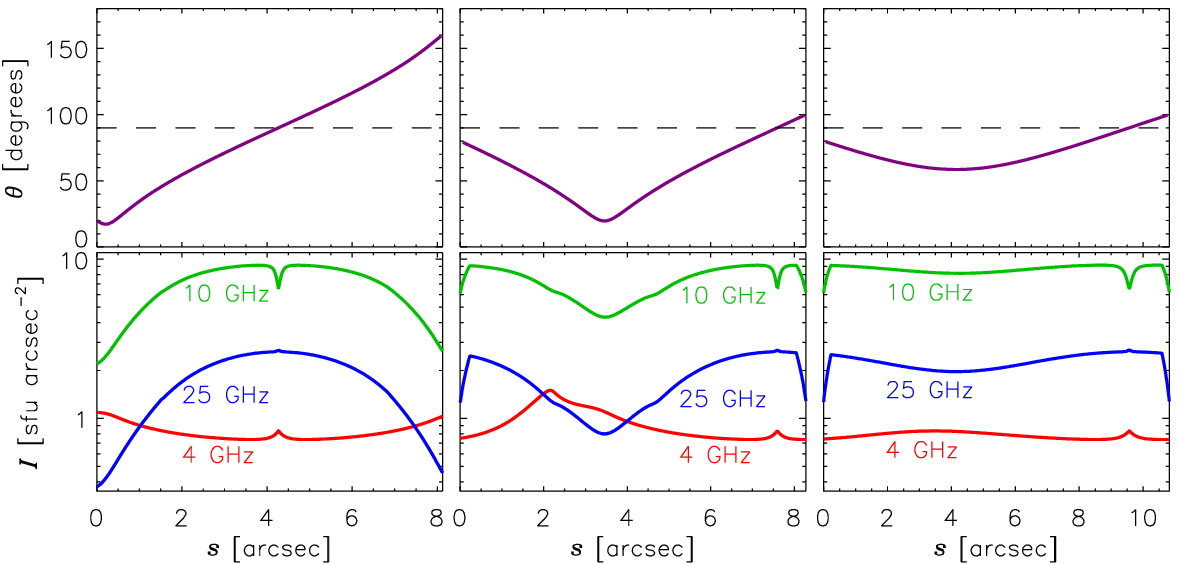}}
\caption{Three orientations of the model magnetic loop considered in the paper: 3D views of the loop (top row); profiles of the viewing angle (relative to the magnetic field, in the equilibrium state) along the loop axis (middle row); simulated profiles of the radio intensity in the equilibrium state at different frequencies (for the low-density model and the electron spectral index of $\delta=4$) along the loop axis (bottom row). The visible coordinate along the loop $s$ is measured from the upper (as shown in the 3D views) footpoint.}
\label{FigLoop}
\end{figure}

As stated before, the aim of this work is to study the effect of the loop curvature on the radio emission. For this, we have created model semi-circular (semi-toroidal) loops by bending the above-mentioned overdense cylinder (see Figure \ref{FigLoop}). In this case, the longitudinal coordinate within the cylinder $z$ is transformed into the distance along the loop axis, and the radial coordinate $r$ -- into the distance from the loop axis (the explicit formulae are given in the Appendix); this allows us to obtain the plasma and magnetic field parameters at any given point. Since the lengths of the solar flaring loops are usually much larger than their radii, we have created the models by stacking several MHD simulation blocks (each corresponding to one wavelength, like in Figure \ref{FigMHD}) along the loop. As the ``basic'' model, we have used the loop containing three wavelengths; such loop has the length of 8475 km and the height (\textit{i.e.}, the curvature radius) of about 2700 km.

The loop orientation is described in general by three Euler angles. However, since rotation around the line of sight does not provide us with any new information (it results simply in the same rotation of the simulated radio images), it is sufficient to consider the variation of only two angles \citep{kuz11}. In this work we assume that the loop is vertical and located at the solar equator. The loop orientation is described by two angles: the angle $\psi$ between the loop plane and the equatorial plane and the longitude $\lambda$. A possible loop inclination is not considered for the same reason as described above: for a vertical loop, the variation of the angles $\psi$ and $\lambda$ is sufficient to obtain any possible loop orientation with respect to the observer (with an accuracy up to rotation around the line of sight); therefore our results are applicable to inclined loops as well. However, we note that the loop locations at the solar disk which are discussed below should be treated as approximate (illustrative), because they correspond to the zero inclination only.

It is not possible to present here the results for all possible orientations; therefore we have focused on three illustrative cases (see Figure \ref{FigLoop} and Table \ref{TabLoops}): 

a) The loop is located near the disk center. In this case, the viewing angle $\theta$ (relative to the magnetic field) varies in a wide range along the loop, being close to zero (or $180^{\circ}$) in the footpoints and being equal to $90^{\circ}$ at some point near the loop top (see Figure \ref{FigLoop}). The coordinate $s$ in Figure \ref{FigLoop} is the visible distance along the loop axis, \textit{i.e.}, it takes into account the projection effects.

b) The loop is located near the solar limb and its plane is almost parallel to the equatorial plane (``Limb A'' orientation). In this case, in contrast to the previous one, the viewing angle is close to $90^{\circ}$ in the loop footpoints and becomes very small near the loop top.

c) The loop is located near the solar limb and its plane is almost perpendicular to the line of sight (``Limb B'' orientation). In this case the viewing angle is always close to $90^{\circ}$ and weakly varies along the loop (although it is not constant, which makes this model different from those studied in the work of \citet{rez14}).

\begin{table}
\caption{Orientations of the magnetic loop used in the simulations: the angle relative to the equatorial plane ($\psi$) and the longitude ($\lambda$).}
\label{TabLoops}
\renewcommand{\tabcolsep}{2.9em}
\begin{tabular}{cccc}
\hline\
Orientation & Disk center & Limb A & Limb B\\
\hline
$\psi$    & $60^{\circ}$ & $20^{\circ}$ & $60^{\circ}$\\
$\lambda$ & $20^{\circ}$ & $80^{\circ}$ & $80^{\circ}$\\
\hline
\end{tabular}
\end{table}

\subsection{Electron Parameters}
We assume that the gyrosynchrotron radio emission is produced by the energetic electrons with the power-law distribution over energy, $f(E)\propto E^{-\delta}$; in the calculations, the energy range is taken to be $\textrm{100 keV}<E<\textrm{10 MeV}$ which is typical for the solar flares. The electrons with lower energies are also present in flares, but their  contribution to the gyrosynchrotron emission is minor (mostly in the form of self-absorption at low frequencies; see, \textit{e.g.}, \citet{hol03}) and usually can be ignored. The pitch-angle distribution of the electrons is assumed to be isotropic (\textit{i.e.}, we assume that it is not affected significantly by the MHD waves). \citet{rez14} have considered the case when the concentration of the energetic electrons $n_{\mathrm{b}}$ inside the cylinder is proportional to the thermal plasma density, $n_{\mathrm{b\,in}}/n_{0\,\mathrm{in}}=\textrm{const}$. Alternatively, one may consider the model in which the energetic electrons are ``frozen'' into the magnetic field, so that their concentration is proportional to the field strength, $n_{\mathrm{b\,in}}/B_{\mathrm{in}}=\textrm{const}$. However, since in the sausage waves the dominant longitudinal component of the magnetic field $B_z$ and the plasma density $n_0$ oscillate in phase, both approaches provide almost identical results. Therefore in the calculations to follow we assume that the concentration of the energetic electrons in the loop is always proportional to the thermal plasma density and $n_{\mathrm{b\,in}}/n_{0\,\mathrm{in}}=0.025$; in particular, in the equilibrium state we have $n_{\mathrm{b\,in}}=10^8$ and $2.5\times 10^8$ $\textrm{cm}^{-3}$ for the low- and high-density models, respectively. We assume that there are no energetic electrons outside the loop ($n_{\mathrm{b\,out}}=0$).

The gyrosynchrotron radio emission is calculated using the Fast Gyrosynchrotron Codes \citep{fle10, kuz11, nit14}; these codes include the contribution of the thermal free-free emission as well. We neglect the possible harmonic structure of the gyrosynchrotron emission at low frequencies by using the ``continuous'' regime of the fast gyrosynchrotron codes, because in real observations this structure should be usually smoothed by the source inhomogeneity \citep{kuz11}; using the ``continuous'' approximation also speeds up the calculations appreciably. The codes calculate the emission characteristics by numerical integration of the radiative transfer equations for the ordinary and extraordinary electromagnetic modes while taking into account the possible mode-coupling effects, \textit{i.e.}, interaction between the modes in the regions of transverse magnetic field \citep{coh60, zhe64}. This approach is sufficient to reproduce accurately the emission intensity in both the quasi-longitudinal and quasi-transverse (with respect to the magnetic field) propagation regimes as well as the circular polarization in the quasi-longitudinal regime \citep{nit14}, although simulating the polarization in the regions with $\theta\simeq 90^{\circ}$ may require considering the full set of Stokes parameters (see the upcoming paper of \citet{rez15}). Only the emission and absorption processes inside the curved magnetic loop are considered, \textit{i.e.}, the emission outside the loop is assumed to propagate like in vacuum.

\begin{figure}
\centerline{\includegraphics{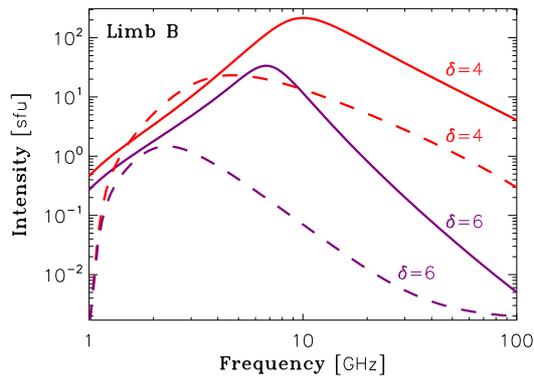}}
\caption{Total (spatially integrated) radio spectra in the equilibrium state. Solid lines: low-density model; dashed lines: high-density model. Two electron spectral indices ($\delta=4$ and 6) and the ``Limb B'' orientation are considered.}
\label{FigSpectrum}
\end{figure}

\section{Results}\label{results}
\subsection{Equilibrium State}
Firstly, we present a brief overview of the emission produced by the loop in the equilibrium state (\textit{i.e.}, without MHD oscillations). This emission is very similar to the emission from an oscillating loop averaged over the wave period. Figure \ref{FigSpectrum} demonstrates the total (spatially integrated) intensity spectra for the loop located at the limb (``Limb B'' orientation). For other orientations, the intensity is slightly lower, but the spectra shapes are qualitatively the same. The concentration of the energetic electrons was chosen to provide that the spectral maximum (for the low-density model and the spectral index of the energetic electrons of $\delta=4$) is located at about 10 GHz -- a typical value for the continuum microwave emission of solar flares.

The bottom row in Figure \ref{FigLoop} demonstrates the simulated distributions of the gyrosynchrotron radio intensity in the equilibrium state along the loop (for the low-density model). The intensity distributions depend on the emission frequency and loop orientation: for the loop located near the disk center, the optically thin emission (\textit{e.g.}, at ${\approx 25}$ GHz) is sharply peaked near the loop top; the same effect is seen for the emission near the spectral peak (at ${\approx 10}$ GHz), while the optically thick emission (\textit{e.g.}, at ${\approx 4}$ GHz) is weakly peaked at the loop footpoints. For the ``Limb A'' orientation, the situation is opposite: now the optically thick emission is peaked near the loop top, while the optically thick emission and the emission near the spectral peak are peaked at the loop footpoints. A similar pattern, although with weaker variations, can be seen in the ``Limb B'' case. In all cases, there are small narrow peaks/dips at the points where the local viewing angle equals $90^{\circ}$; they are caused by different polarization characteristics of the ordinary and extraordinary electromagnetic modes and are typical of the gyrosynchrotron radiation.

\subsection{Intensity Modulation: 2D Images}
Oscillation-induced variations of the radio emission in different parts of the magnetic loop tend to compensate each other, thus reducing the variations of the total (spatially unresolved) emission. Due to the variation of the viewing angle (and hence of the local emission parameters) along the loop, the mentioned compensation is not complete. Nevertheless, the variations of the total intensity are always much weaker than those of the local (spatially resolved) emission; amplitude of the total intensity variations decreases with a decrease of the relative wavelength of the MHD waves (\textit{i.e.}, with an increase of the number of the waves in the loop). Therefore we focus below on the characteristics of the spatially-resolved emission.

\begin{figure}
\centerline{\includegraphics{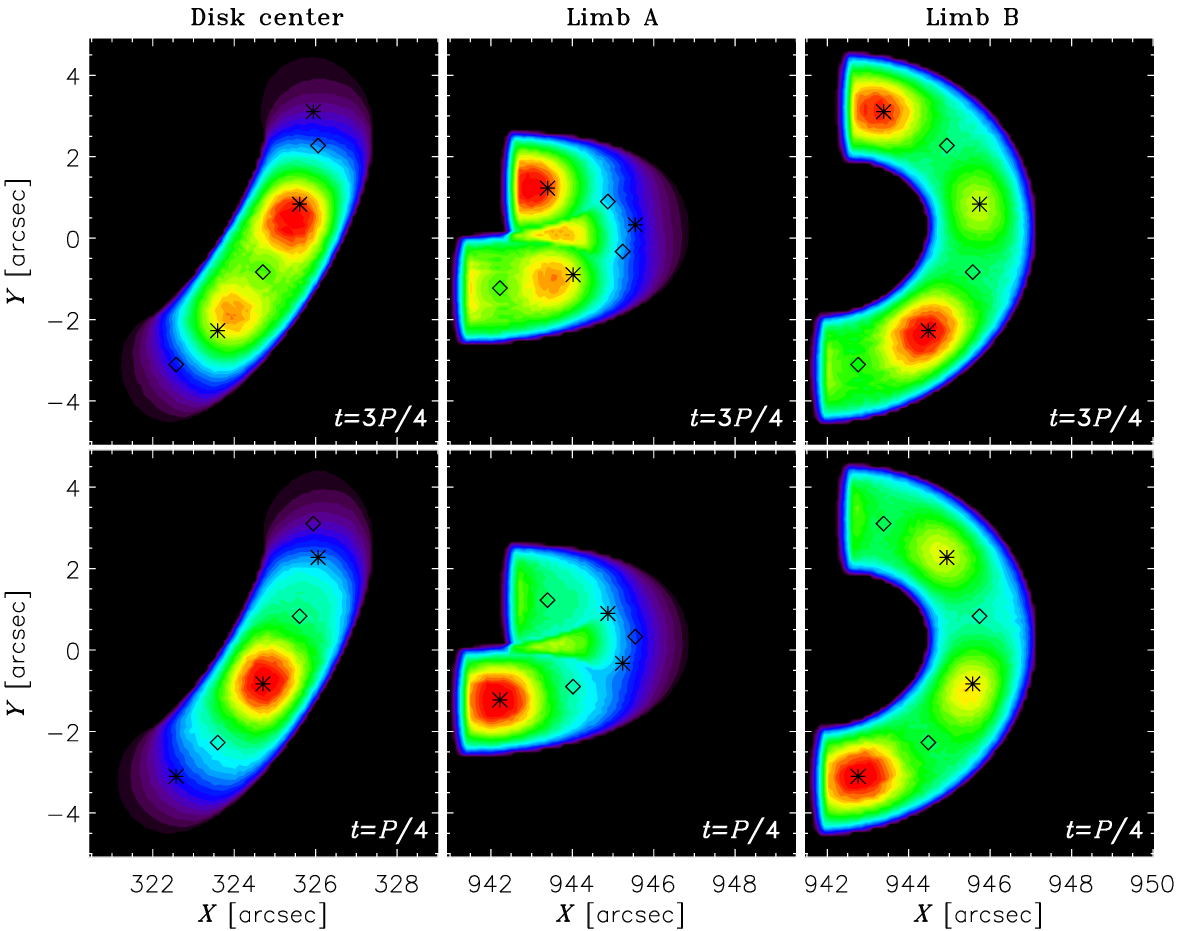}}
\flushright{\includegraphics{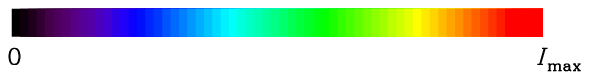}}
\caption{Two-dimensional images of the microwave emission (at the frequency of 25 GHz) for the standing wave. The low-density model is used, the loop contains three sausage waves and the electron spectral index is $\delta=4$. Two oscillation phases corresponding to the maximum deviations from the equilibrium are shown. The asterisks ($\ast$) and diamonds ($\diamond$) mark the points with the maximum / minimum magnetic field at the loop axis.}
\label{FigImages}
\end{figure}

Figure \ref{FigImages} demonstrates the simulated radio images at the frequency of 25 GHz (optically thin emission) for the standing wave. Two opposite oscillation phases are shown. To make the effect of the MHD waves more pronounced, the intensity variations were enhanced as 
\begin{equation}
I\to\left<I\right>+5(I-\left<I\right>), 
\end{equation}
where $\left<I\right>$ is the intensity averaged over the wave period $P$:
\begin{equation}
\left<I\right>=\frac{1}{P}\int_0^PI(t)\,\mathrm{d}t.
\end{equation}
We can see in the figure that for the ``Limb B'' orientation (with an almost constant viewing angle) the visible emission structure, in general, reflects the structure of the MHD wave: there are three intensity maxima corresponding to the points of maximum magnetic field (and hence the maximum concentration of energetic electrons). For other orientations, due to the projection effects, the intensity maxima/minima slightly deviate from the magnetic field maxima/minima; also, not all MHD waves are visible (\textit{i.e.}, both the intensity and the oscillation-induced variation amplitude depend strongly on the coordinate along the loop). For the propagating MHD waves, the radio images are similar to those shown in Figure \ref{FigImages}, but the brighter/darker spots drift gradually along the loop with time.

\subsection{Phase Relations and Modulation Depth}\label{phase}
We are interested primarily in the qualitative effects that could reveal the basic MHD wave characteristics (such as the wave mode). As mentioned above, in real flaring loops the MHD wave properties (including the phase speed) can vary strongly along the loop, which is not considered in our model. Therefore it makes no sense to compare, \textit{e.g.}, the simulated time profiles of the emission at the loop top and the footpoints. Instead we consider the local emission parameters, \textit{i.e.}, the temporal delays (phase differences) between the emissions at different frequencies at a given point.

\begin{figure}
\centerline{\includegraphics{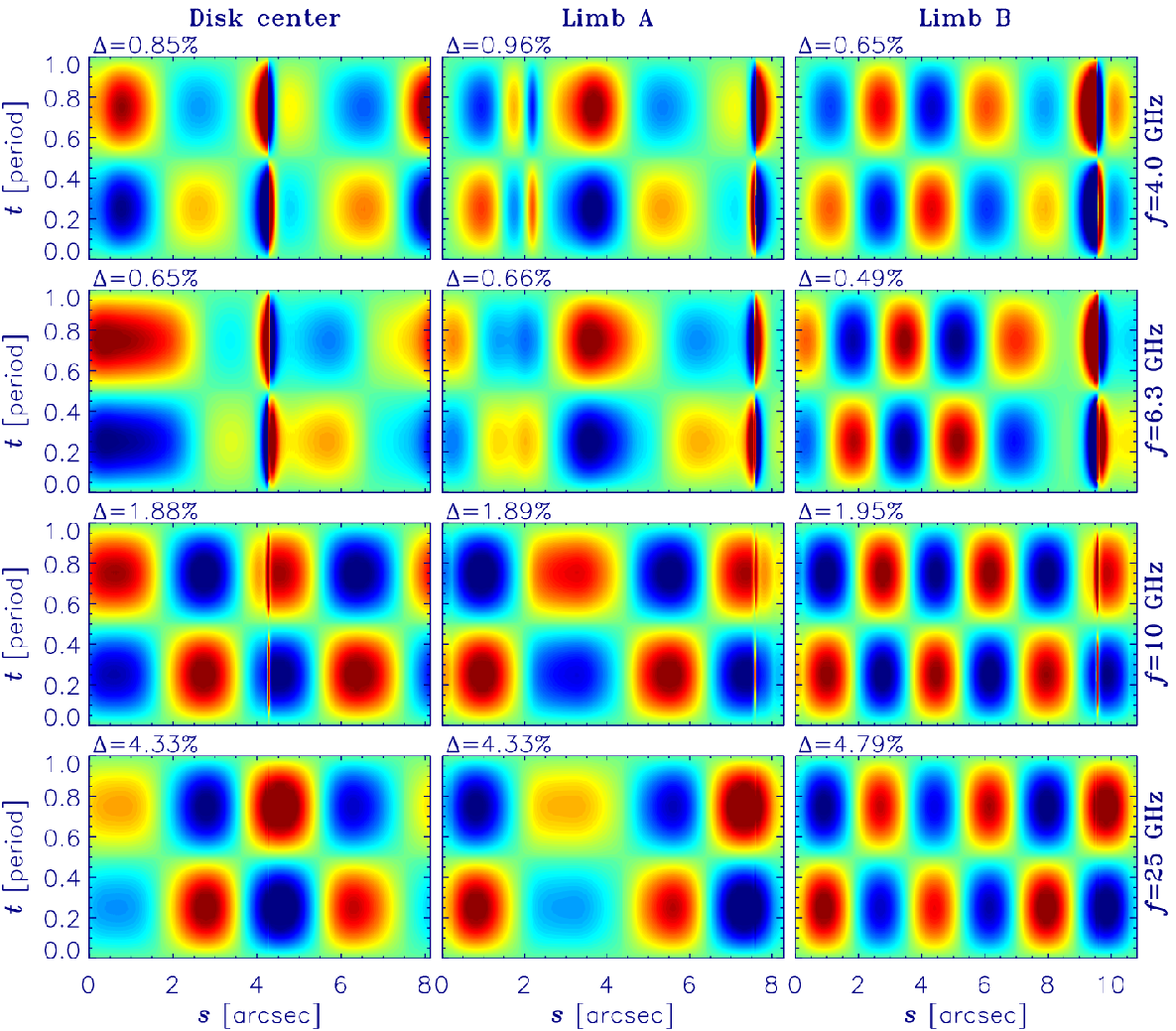}}
\flushright{\includegraphics{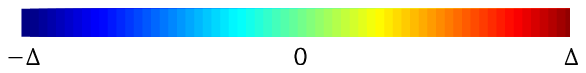}}
\caption{Time-distance plots of the intensity variation: the normalized intensity variation $(I-\left<I\right>)/\left<I\right>$ at the loop axis \textit{vs.} the visible coordinate along the loop axis and the oscillation phase. The low-density model (standing wave) is used; the loop contains three sausage waves and the electron spectral index is $\delta=4$. For each emission frequency and loop orientation, the maximum modulation depth $\Delta$ is given above the corresponding panel.}
\label{FigTDs}
\end{figure}
\begin{figure}
\centerline{\includegraphics{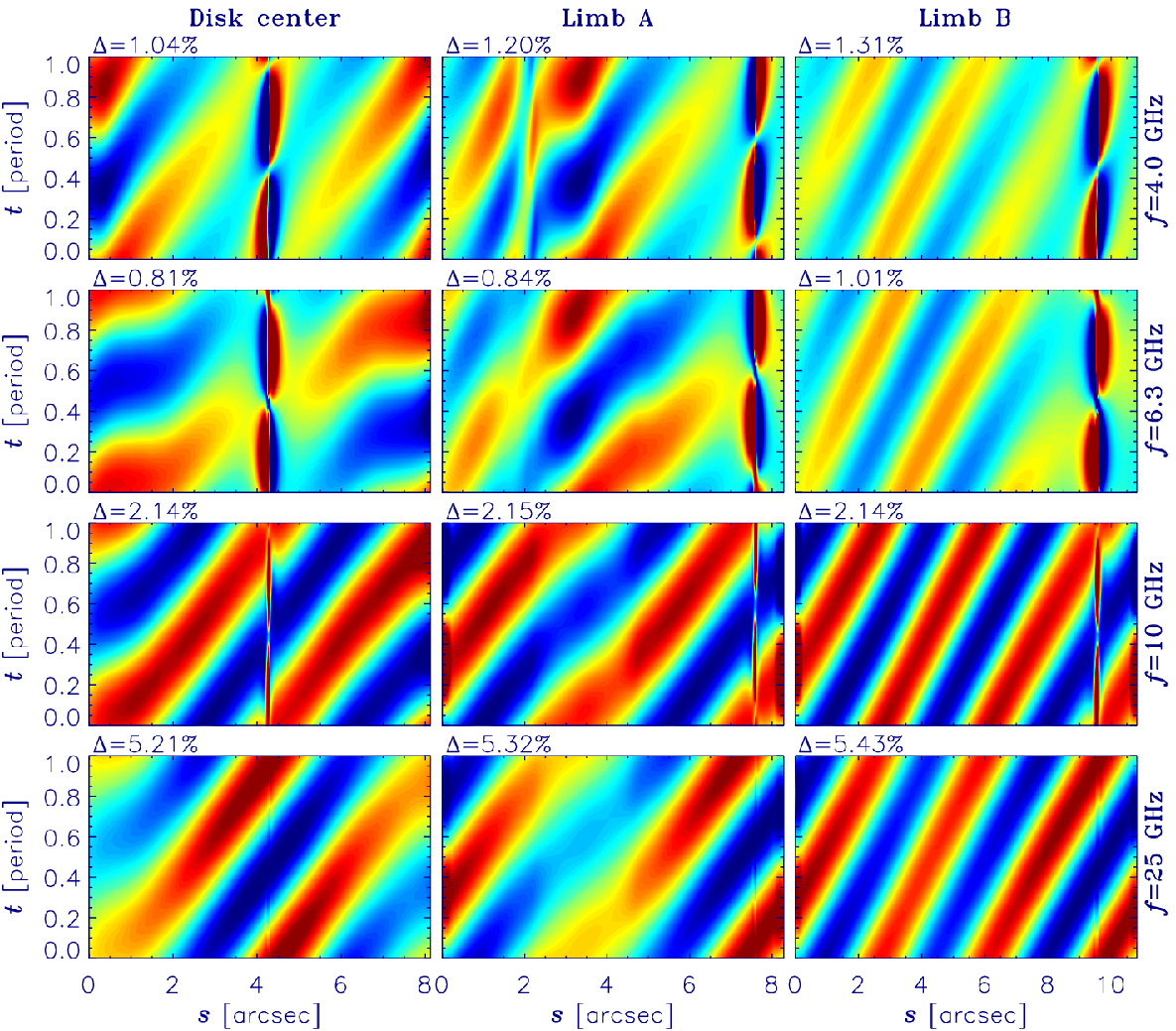}}
\flushright{\includegraphics{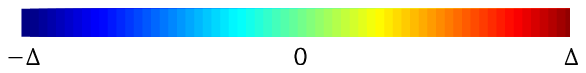}}
\caption{Same as in Figure \protect\ref{FigTDs}, for the propagating wave.}
\label{FigTDp}
\end{figure}

Figures \ref{FigTDs}-\ref{FigTDp} show the time-distance plots of the normalized intensity variation $(I-\left<I\right>)/\left<I\right>$ at the axis of the magnetic loop (\textit{i.e.}, along the line formed by the projection of the loop axis on the image plane) for the low-density model. Four frequencies are shown, representing the optically thick case (4 GHz), ``moderately'' optically thick case ($6.3$ GHz), spectral peak (10 GHz), and optically thin case (25 GHz). Above each panel, the respective maximum modulation depth $\Delta$ is given; this value is defined as
\begin{equation}
\Delta=\max\left|\frac{I-\left<I\right>}{\left<I\right>}\right|.
\end{equation} 
Note that the maximum modulation depth $\Delta$ is different from the average modulation depth $\epsilon$ given by Equation (\ref{epsilon}) (which will be used in Section \ref{index}). According to Figures \ref{FigTDs}-\ref{FigTDp}, the maximum modulation depth $\Delta$ varies from ${\lesssim 1\%}$ at low frequencies to ${\gtrsim 5\%}$ at high frequencies. These values are relatively low and, in fact, it would not be easy to detect such oscillations in real observations. However, this is because we consider the relatively low-amplitude MHD waves: the variations of the emission source parameters $\delta B/\left<B\right>$ and $\delta n_{\mathrm{b}}/\left<n_{\mathrm{b}}\right>$ do not exceed $2.5\%$ and thus the variations of the intensity are comparable to the MHD wave amplitude. The quasi-periodic pulsations reported by \citet{nak03}, \citet{mel05}, and \citet{rez07} (with the modulation depths of up to ${\approx 0.3}$) require much more powerful MHD waves; nevertheless, if those pulsations are caused by sausage MHD waves, their characteristics should be qualitatively similar to the case studied in this work. Note also that the intensity variations are more pronounced in the optically thin frequency range, in agreement with the results of \citet{mos12} and \citet{rez14}. 

Figure \ref{FigTDs} demonstrates the simulation results for the standing wave. We can see that the relative modulation depth, in general (excluding the regions with $\theta\simeq 90^{\circ}$), follows the same trends as the unperturbed intensity (see Figure \ref{FigLoop}): for the loop near the disk center, the modulation depth increases towards the loop top for the optically thin emission (${\gtrsim 10}$ GHz) and increases towards the loop footpoints for the optically thick emission (${<10}$ GHz); for the loops near the limb, the trends are opposite. Oscillations at low ($\lesssim 4$ GHz, the first row of the figure) and high (${\gtrsim 10}$ GHz, third and fourth rows) frequencies are synchronous. The fact that the optically thin and thick emissions oscillate in phase agrees with the results of \citet{rez14}, but contradicts the conclusions of \citet{mos12}. The reason is that, in contrast to \citet{mos12}, we take into account the modulation of the viewing angle by the sausage wave but neglect the modulation of the energy of the nonthermal electrons. In addition, we use the spatially inhomogeneous model in which the source parameters vary along the line of sight; this is especially important for the optically thick emission. Since two different approaches (ours and that of \citet{mos12}) provide opposite results, we conclude that the sausage-mode MHD waves (at least, in application to the radio intensity modulation) cannot be treated without a spatially resolved 3D model. Note that this behaviour (in-phase variations of the optically thin and thick emissions) is observed for all loop orientations and at all locations along the loop; \textit{i.e.}, it is independent of the viewing angle (except the $\theta=90^{\circ}$ case which is discussed below). 

An interesting feature is observed at ``intermediate'' frequencies (${\approx 6}$ GHz, the second row of Figure \ref{FigTDs}), \textit{i.e.}, in the optically thick range, but not far from the spectral peak; the optical depth $\tau$ here exceeds unity, but is not very large (${\tau\approx 10}$). We can see that the intensity oscillations at these frequencies are shifted by about $1/4$ wavelength with respect to the oscillations at higher and lower frequencies. This is caused by the wave-induced variations of the viewing angle: at the intermediate frequencies, these variations (albeit very small) become the main factor modulating the intensity. In contrast, at higher or lower frequencies (with ${\tau<1}$ or ${\tau\gg 1}$), the effect of changing the magnetic field and electron concentration dominates. At the same time, the modulation depth at the intermediate frequencies reaches its minimum (in comparison with other frequency ranges). 

The oscillating magnetic field direction (and hence the oscillating viewing angle) also produces some peculiarities in the parts of the loop where the viewing angle is close to $90^{\circ}$. Changing the sign of the magnetic field projection on the line of sight results in changing the dominating electromagnetic wave mode which affects the intensity. As a result, in the optically thick frequency range the MHD wave induces intensity variations of opposite signs at different sides of the $\theta=90^{\circ}$ stripe; the modulation depth increases sharply. In the optically thin frequency range the intensity variations at different sides of the $\theta=90^{\circ}$ stripe have the same sign and the modulation depth demonstrates only minor variations. It should be noted, however, that the mentioned peculiarities occur only in very narrow stripes and thus their detection in real observations requires very high spatial resolution. 

Similar peculiarities (although without significant changes of the modulation depth) can be seen for the ``Limb A'' orientation, in the optically thick frequency range, at the distance of about $s\simeq 2$'' from the upper footpoint. They occur due to the specific loop topology: the optically thick emission is produced near the loop boundary and, for the ``Limb A'' orientation (see Figure \ref{FigLoop}), the dependence of the viewing angle at the loop boundary surface on the coordinate $s$ can become discontinuous; for other loop orientations, this effect is not observed.

Figure \ref{FigTDp} demonstrates the simulation results for the propagating wave. Since the structure of the standing and propagating sausage waves is the same, all the above conclusions about the radio intensity oscillations remain valid; the maximum modulation depths shown in Figure \ref{FigTDp} are slightly higher than in Figure \ref{FigTDs}. Using Figure \ref{FigTDp}, we can track propagation of the MHD waves and estimate their speed. For the low-density model considered here (see Table \ref{TabMHD}), the phase speed of the sausage waves equals 8070 km $\textrm{s}^{-1}$; the visible propagation speed of the radio brightenings/darkenings for the ``Limb B'' orientation is very close to this value. However, due to the varying viewing angle, the projected (visible) wave speed varies along the loop; this effect is especially important for the ``Disk center'' and ``Limb A'' orientations. The amplitude of the propagating radio intensity variations also varies along the loop with the most notable effects occurring near the regions of transverse (with respect to the line of sight) magnetic field: in these regions, the bright/dark stripes in Figure \ref{FigTDp} become discontinuous, although the MHD wave itself has no peculiarities; the bright/dark stripes in the time-distance plots (in the optically thick frequency range) can also become discontinuous in the regions where the viewing angle at the loop boundary sharply depends on the coordinates in the image plane. Thus when interpreting the radio observations with quasi-periodic pulsations, a special attention should be paid to the $\theta=90^{\circ}$ layers and/or the regions where the visible loop boundary demonstrates topological peculiarities, since they can introduce additional phase shifts which do not actually reflect the properties of the underlying MHD waves. 

Assuming that the relative intensity variations are approximately proportional to the relative MHD wave amplitude, we can estimate that the MHD waves with the amplitude $\delta B/\left<B\right>$ of about 40-50\% (with other source parameters being the same as in Figures \ref{FigTDs}-\ref{FigTDp}) should produce the intensity modulation depth of up to 100\%. However, simulating the MHD waves with such amplitudes requires taking into account the nonlinear effects. As will be shown in Sections \ref{index}-\ref{razin}, the emission modulation depth becomes higher for softer electron beams or for higher plasma densities at low frequencies (when the Razin effect is important), although this increase of the modulation depth is always accompanied by a sharp decrease of the intensity.

\subsection{Parametric Study}
We now consider how the fluctuations of the radio emission depend on various parameters of the flaring loop.

\begin{figure}
\centerline{\includegraphics{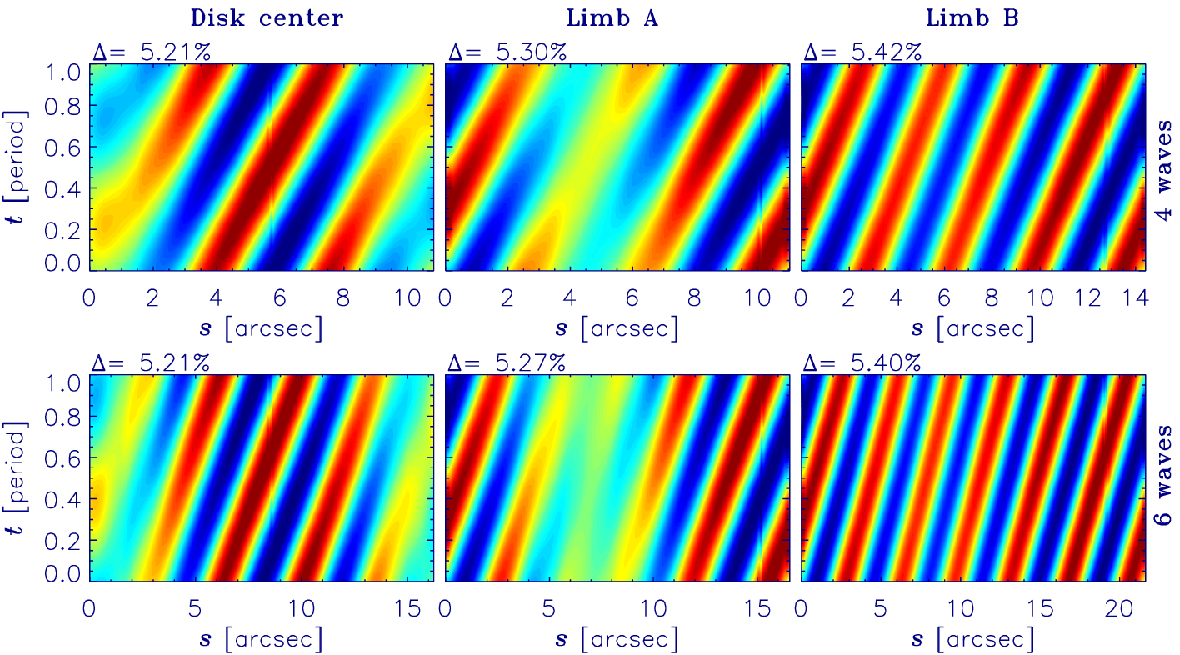}}
\flushright{\includegraphics{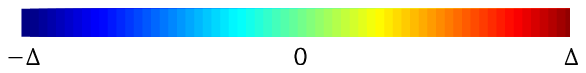}}
\caption{Time-distance plots of the intensity variation at the loop axis (same as in Figures \protect\ref{FigTDs}-\protect\ref{FigTDp}) at the frequency of 25 GHz for different loop orientations and different numbers of wavelengths (four and six) in the loop. The low-density model (propagating wave) is used and the electron spectral index is $\delta=4$.}
\label{FigWp}
\end{figure}

\subsubsection{Effect of the Wave Number}
Figure \ref{FigWp} demonstrates the time-distance plots of the intensity variation for the models with increased numbers of MHD waves in the magnetic loop (four and six waves, which corresponds to the loop lengths of 11\,300 and 16\,950 km, respectively); the simulation results shown correspond to the low-density model, nonthermal electrons spectral index of $\delta=4$, and the optically thin emission (at the frequency of 25 GHz). As already shown above, the standing and propagating MHD waves affect the radio emission in a qualitatively similar way; therefore only the results for the propagating wave are shown.

Our simulations have revealed no significant effect of the number of the MHD waves in the loop, \textit{i.e.}, both the emission modulation depth and the phase relations between the emissions at different frequencies remain the same. The only difference can be noticed in the regions where the loop axis becomes nearly parallel to the line of sight, \textit{i.e.}, near the loop footpoints for the ``Disk center'' orientation and near the loop top for the ``Limb A'' orientation: since the length of the formation region of the optically thin emission can exceed an MHD wavelength, the effects of variation of the magnetic field and electron concentration in different parts of the emission source (with opposite phases) can compensate each other. As a result, the amplitude of the intensity oscillations decreases with a decrease of the relative (with respect to the loop length) MHD wavelength. This effect is not observed in the optically thick frequency range.

It can be seen in Figure \ref{FigWp} (as well as in Figures \ref{FigImages}-\ref{FigTDp}) that the MHD observations are best resolved when the magnetic tube is observed in a nearly perpendicular direction. Indeed, for the ``Limb B'' orientation we can see one-to-one correspondence between the MHD waves and the spatial and temporal structure of the radio emission. For other orientations, this correspondence is less straightforward and some MHD wave nodes can even be missing in the radio emission; as expected, the radio emission cannot accurately reflect the MHD wave structure in the regions where the viewing angle with respect to the magnetic field $\theta$ is small. While it is not easy to estimate the number of observed MHD wave nodes in a general case, we can expect (and this is confirmed by the simulations) that the MHD wave nodes become missing in the radio emission in the regions with $\tan\theta\lesssim R/\lambda$, where $R$ is the loop radius and $\lambda$ is the MHD wavelength; this happens either due to the averaging along the line of sight (in the optically thin frequency range) or due to the strong reabsorption of the emission during propagation (in the optically thick frequency range). Note also that, as has been discussed above, the $\theta=90^{\circ}$ layers can introduce the patterns in the radio emission (at low frequencies) similar to additional MHD waves.

\subsubsection{Effect of the Spectral Index of Energetic Electrons}\label{index}
The energy spectrum of the radio-emitting electrons affects greatly the gyrosynchrotron emission. As can be seen from Figure \ref{FigSpectrum}, softer electron beams (with larger spectral indices $\delta$) produce less intense emission with steeper spectrum slope in the optically thin frequency range; the spectral peak shifts to lower frequencies. We have found that variation of the spectral index $\delta$ does not change the simulated radio images (like in Figure \ref{FigImages}) qualitatively. However, there are quantitative changes in the emission time profiles which may affect detectability of the MHD oscillations.

\begin{figure}
\centerline{\includegraphics{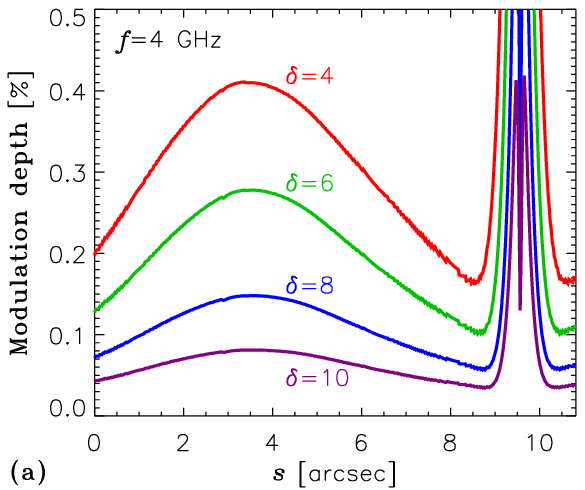}
\includegraphics{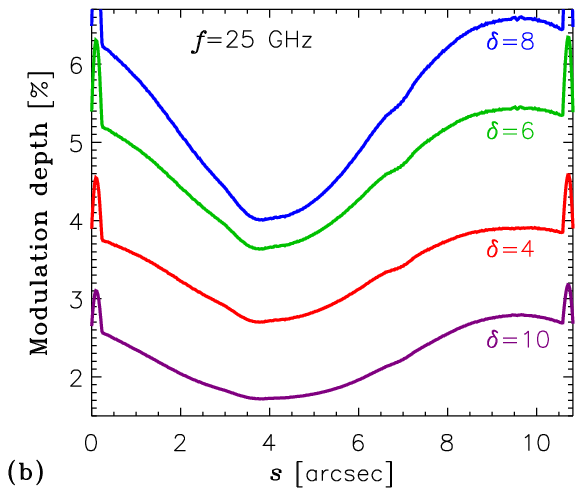}}
\caption{Average modulation depth of the intensity at the loop axis for different frequencies and different spectral indices of the energetic electrons. The low-density model (propagating wave) and the ``Limb B'' orientation are considered.}
\label{FigIndex}
\end{figure}

Figure \ref{FigIndex} demonstrates the average modulation depths of the intensity (for the propagating MHD wave, three sausage waves per loop, low-density model, and ``Limb B'' orientation); the emission at the loop axis is considered and the values are plotted as functions of the coordinate along the loop. The average modulation depth $\epsilon$ is calculated as \citep{mos12}
\begin{equation}\label{epsilon}
\epsilon=\frac{m}{\left<I\right>},\qquad
m^2=\frac{1}{P}\int_0^P\left[I(t)-\left<I\right>\right]^2\mathrm{d}t;
\end{equation}
it is related to the maximum modulation depth $\Delta$ (shown in Figures \ref{FigTDs}-\ref{FigWp}) as $\Delta\simeq\epsilon\sqrt{2}$. We can see that in the optically thin frequency range (Figure \ref{FigIndex}b) an increase in the spectral index results firstly in an increase in the modulation depth. As a result, the oscillation-modulated radio images become more contrasted and the periodic variations of the emission become easier to detect. For even higher electron spectral indices (${\delta\gtrsim 10}$ for the considered frequency of 25 GHz), the modulation depth decreases sharply, because the gyrosynchrotron emission becomes too weak and the emission is produced mainly by the free-free mechanism that is weakly affected by the MHD oscillations. In the optically thick frequency range (Figure \ref{FigIndex}a) the modulation depths are low (${\lesssim 0.5\%}$) and decrease with increasing $\delta$.

As mentioned above, Figure \ref{FigIndex} corresponds to the propagating wave and ``Limb B'' orientation; for other orientations and/or for the standing wave, the above conclusions remain the same. Assuming that the relative intensity variations are approximately proportional to the relative MHD wave amplitude, we can estimate that for the electron spectral index of $\delta=8$ and the MHD wave amplitude $\delta B/\left<B\right>$ of about 25-30\%, the average ($\epsilon$) and maximum ($\Delta$) modulation depths in the optically thin frequency range should reach ${\approx 70\%}$ and ${\approx 100\%}$, respectively; the MHD waves with such amplitudes seem to be feasible.

\subsubsection{Effect of the Plasma Density}\label{razin}
Finally, we consider the case when the spectral peak of the gyrosynchrotron radio emission is formed by the Razin effect \citep{raz60a, raz60b}. This effect takes place in a relatively high-density plasma and results in a considerable suppression of the gyrosynchrotron emission at the frequencies below
\begin{equation}
f_{\mathrm{R}}\simeq\frac{2}{3}\frac{f_{\mathrm{p}}^2}{f_{\mathrm{B}}},
\end{equation}
where $f_{\mathrm{p}}$ and $f_{\mathrm{B}}$ are the plasma and cyclotron frequencies in the emission source, respectively. At the frequencies below $f_{\mathrm{R}}$, the variations of the thermal plasma density should become the main factor affecting the radio intensity.

As an illustrative example, we use the high-density model (see Table \ref{TabMHD}). As can be seen from Figure \ref{FigSpectrum}, in this case the Razin frequency $f_{\mathrm{R}}$ is about 4 GHz. Thus at the frequencies well above 4 GHz the intensity variations should be caused mainly by the oscillations of the magnetic field and energetic electron concentration (like in the low-density model), at the frequencies below 4 GHz the intensity variations should be caused mainly by the Razin effect (\textit{i.e.}, by the oscillating plasma density), while at the frequencies around the spectral peak (${\approx 4}$ GHz) both factors should be important.

\begin{figure}
\centerline{\includegraphics{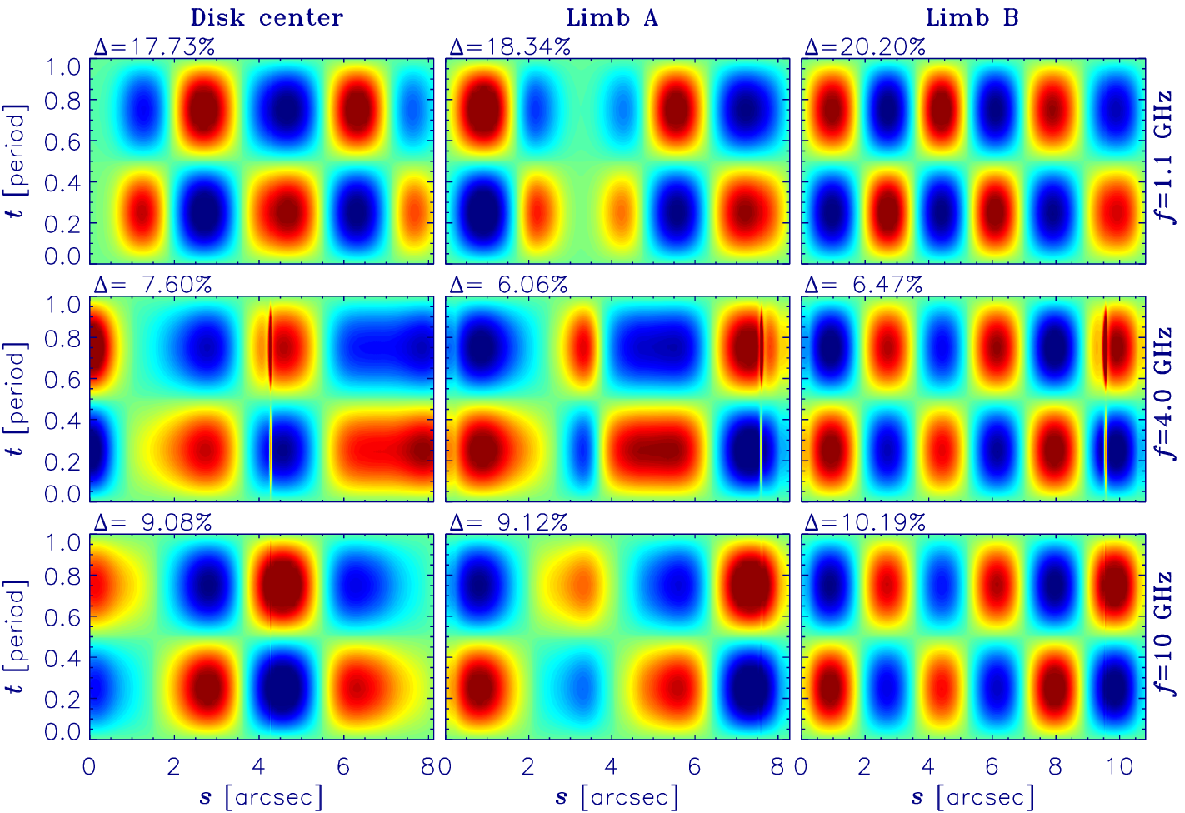}}
\flushright{\includegraphics{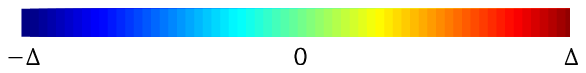}}
\caption{Time-distance plots of the intensity variation at the loop axis (same as in Figures \protect\ref{FigTDs}--\protect\ref{FigWp}) for the high-density model. Standing wave is considered and the electron spectral index is $\delta=4$.}
\label{FigHD}
\end{figure}

Figure \ref{FigHD} demonstrates the simulation results (time-distance plots of the intensity variation) at three different frequencies for the high-density model, standing sausage wave, and the nonthermal electrons spectral index of $\delta=4$. In the optically thin frequency range, the modulation depth of the intensity is larger than that for the low-density model due to the model parameters used (the relative MHD wave amplitude for the high-density model is larger, see Table \ref{TabMHD}). The modulation depth reaches its minimum at a frequency below the spectral peak (${\approx 2}$ GHz, depending on the viewing angle) and then rapidly increases with decreasing frequency; however, as can be seen in Figure \ref{FigSpectrum}, the intensity at those frequencies also becomes very low, so that detection of the intensity oscillations in observations would be difficult.

It can be seen in Figure \ref{FigHD} that for the ``Limb B'' orientation (when the viewing angle is close to $90^{\circ}$) the radio oscillations in the optically thin frequency range (${\gtrsim 10}$ GHz) and near the spectral peak (${\approx 4}$ GHz) are synchronous (in phase), while the low-frequency emission (${\lesssim 2}$ GHz) oscillates in anti-phase with the emissions at higher frequencies; the phase reversal occurs (for ${\theta\simeq 90^{\circ}}$) at ${\approx 2}$ GHz. Clearly, this result indicates that at high frequencies (above and even slightly below the spectral peak) the intensity variations are caused by the variations of the magnetic field and nonthermal electrons concentration, while at lower frequencies (${<2}$ GHz) the Razin effect dominates; these factors affect the emission in opposite ways.

A similar picture can be seen for two other loop orientations in the regions where the viewing angle is close to $90^{\circ}$ (\textit{i.e.}, near the loop top for the ``Disk center'' orientation and near the footpoints for the ``Limb A'' orientation). For smaller viewing angles the picture is more complicated, since both the modulation depth and the above-mentioned frequency of the oscillation phase reversal are angle-dependent: the frequency of the phase reversal increases with a decrease of the viewing angle. As a result, if the viewing angle varies gradually along the loop, we can see either correlation or anti-correlation between the oscillations at two given frequencies, depending on the coordinate $s$.

The above results contradict the conclusions of \citet{mos12} who have found that the oscillations of the high- and low-frequency emissions (in the case when the Razin effect is important) should occur strictly in phase. As mentioned before (see Section \ref{phase}), this is because we consider a slightly different model for the variation of the parameters in the sausage wave, as well as the spatially inhomogeneous emission source. Our results also disagree with the conclusions of \citet{rez14} who have found that the oscillations of the high- and low-frequency emissions occur mostly in phase (although this conclusion was dependent on the viewing angle). Most likely, this is because \citet{rez14} considered the emission with the frequencies below the spectral peak but above the ``phase reversal frequency'' so that the effect of the Razin suppression was too weak.

\section{Conclusions}\label{conclusion}
We have simulated the modulation of the gyrosynchrotron radio emission by the sausage-mode MHD oscillations in a semi-circular magnetic loop, in application to the solar flares. Although the model used does not include the variation of the magnetic field and plasma parameters with height, it has allowed us to study the effect of varying the viewing angle along the loop. We have considered three loop orientations and different parameters of the magnetic field, plasma, and energetic particles. In particular, we have studied in detail the phase relations between the spatially resolved intensity oscillations at different frequencies. The results can be summarized as:
\begin{itemize}
\item
In the model with a relatively low plasma density (when the Razin effect is negligible), the high-frequency (optically thin) and low-frequency (optically thick) emissions oscillate, as a rule, in phase. An exception is the optically thick emission at the frequencies just below the spectral peak, whose oscillations are shifted by ${\approx 1/4}$ MHD wavelength with respect to the oscillations at other frequencies.
\item
In the model with a relatively high plasma density (when the Razin effect is important), the emissions at high and low frequencies oscillate in anti-phase. The reversal of the oscillation phase occurs, as a rule, at a frequency well below the spectral peak; this frequency depends on the viewing angle.
\item
In the optically thin frequency range, both the unperturbed intensity and the absolute and relative intensity variations are the highest in the regions where the local magnetic field is nearly perpendicular to the line of sight. In contrast, in the optically thick frequency range all the mentioned values are the highest in the regions with small viewing angles.
\item
In the optically thin frequency range, the intensity oscillations are more pronounced for softer electron beams (\textit{i.e.}, with higher spectral indices). In the optically thick frequency range the trend is opposite.
\item
In the optically thick frequency range, the intensity oscillations can have sharp phase shifts between the adjacent regions due to the presence of layers of transverse (with respect to the line of sight) magnetic field and other topological peculiarities.
\item
Averaging over the visible source area and/or along the line of sight reduces the amplitude of the intensity oscillations; this effect becomes more important with an increase in the number of the MHD harmonics in the loop.
\end{itemize}
The latter result means that, due to a limited spatial resolution, the real observations can differ from the ideal simulations presented here. Nevertheless, we expect that if an instrument spatial resolution is better than a MHD wavelength, this will allow us to detect (or rule out) the above-mentioned phase relations between the oscillations at different frequencies; another requirement is the ability to perform imaging observations at many frequencies. We anticipate that the forthcoming multiwavelength radioheliographs (such as the Expanded Owens Valley Solar Array, Upgraded Siberian Solar Radio Telescope, and Chinese Spectral Radioheliograph; see the review of \citet{nak14} and references therein) will be able to provide the necessary data.

\begin{acks}
This work was supported in part by the Russian Foundation of Basic Research (grants 12-02-00173, 13-02-90472 and 14-02-91157) and by a Marie Curie International Research Staff Exchange Scheme ``Radiosun'' (PEOPLE-2011-IRSES-295272). The research leading to the presented results has been sponsored by an Odysseus grant of the FWO Vlaanderen. It has been performed in the context of the IAP P7/08 CHARM (Belspo) and the GOA-2015-014 (KU Leuven).
\end{acks}

\appendix
\section{Toroidal Coordinates}\label{toroidal}
\begin{figure}
\centerline{\includegraphics{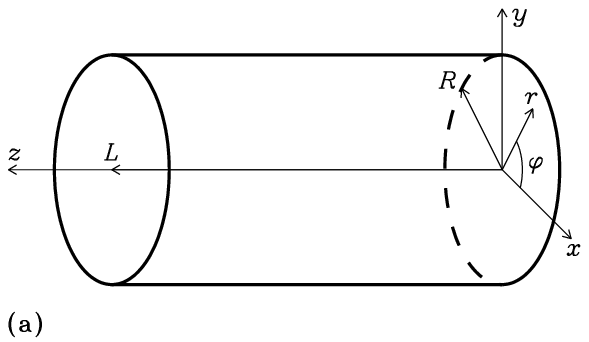}
\includegraphics{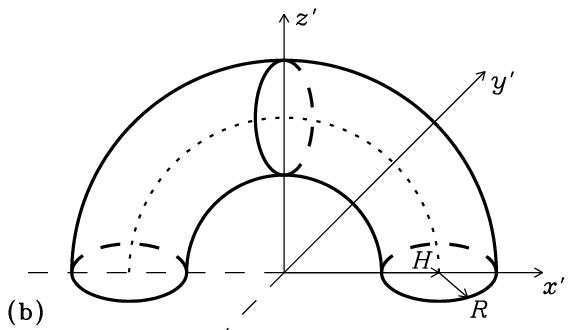}}
\caption{a) Sketch of the model plasma cylinder. b) Sketch of the model semi-toroidal loop.}
\label{FigCoordinates}
\end{figure}

As was mentioned above, we consider the MHD waves in an overdense cylinder; the system can be described by the cylindrical coordinates: the distance along the cylinder axis $z$, the distance from this axis $r$, and the azimuthal angle $\varphi$ (see Figure \ref{FigCoordinates}a). We assume that the cylinder has the length $L$ (\textit{i.e.}, $0\le z\le L$) and the radius $R$ (the radio emission is produced only inside the cylinder, \textit{i.e.}, at $r\le R$).

The cylinder is assumed to be bent into a semi-circular (semi-toroidal) loop (see Figure \ref{FigCoordinates}b); the loop curvature radius is $H=L/\pi$ (evidently, we need $H>R$). We assume that initially the loop is oriented as shown in Figure \ref{FigCoordinates}b, \textit{i.e.}, it is vertical, symmetric with respect to the $x'z'$ and $y'z'$ planes, and its axial line lies in the $x'z'$ plane. Then the cylindrical coordinates $(z, r, \varphi)$ are related to the Cartesian coordinates $(x', y', z')$ as
\begin{eqnarray}
x'&=&(r\cos\varphi+H)\cos\chi,\\
y'&=&r\sin\varphi,\\
z'&=&(r\cos\varphi+H)\sin\chi,
\end{eqnarray}
and
\begin{eqnarray}
\tan\chi&=&\frac{z'}{x'},\\
r^2&=&(x'-H\cos\chi)^2+(y')^2+(z'-H\sin\chi)^2,\\
\tan\varphi&=&\frac{y'}{x'\cos\chi+z'\sin\chi-H},
\end{eqnarray}
where $\chi=z/L$.

Assume that a vector $\mathbf{A}$ at a given point is defined by its three components $(A_z, A_r, A_{\varphi})$ in the cylindrical coordinate system. After transformation to the semi-circular loop, the vector can be expressed as $\mathbf{A}'=A_z\mathbf{a}'_z+A_r\mathbf{a}'_r+A_{\varphi}\mathbf{a}'_{\varphi}$, where the unit vectors $\mathbf{a}'_{z, r, \varphi}$ in the Cartesian coordinates are given by
\begin{eqnarray}
\mathbf{a}'_z&=&(-\sin\chi, 0, \cos\chi),\\
\mathbf{a}'_r&=&(\cos\varphi\cos\chi, \sin\varphi, \cos\varphi\sin\chi),\\
\mathbf{a}'_{\varphi}&=&(-\sin\varphi\cos\chi, \cos\varphi, -\sin\varphi\sin\chi).
\end{eqnarray}
The above formulae are valid in a general case. In this work we consider the azimuthally symmetric case when all parameters are independent of the azimuthal angle $\varphi$ and the azimuthal components $A_{\varphi}$ of all vectors are equal to zero.

The orientation shown in Figure \ref{FigCoordinates}b corresponds to the vertical loop located at the center of the solar disk, with the axis $z'$ being the line of sight and the plane $x'z'$ being the equatorial plane. All other orientations (see, \textit{e.g.}, Figure \ref{FigLoop}) are obtained by rotating the loop firstly by the angle $\psi$ around the local vertical ($z'$ axis) and then by the angle $\lambda$ around the solar rotation axis.

\end{article}

\begin{thebibliography}{59}
\ifx\bisbn     \undefined \def\bisbn  #1{ISBN #1}\fi
\ifx\binits    \undefined \def\binits#1{#1}\fi
\ifx\bauthor   \undefined \def\bauthor#1{#1}\fi
\ifx\batitle   \undefined \def\batitle#1{#1}\fi
\ifx\bjtitle   \undefined \def\bjtitle#1{\textit{#1}}\fi
\ifx\bvolume   \undefined \def\bvolume#1{\textbf{#1}}\fi
\ifx\byear     \undefined \def\byear#1{#1}\fi
\ifx\bissue    \undefined \def\bissue#1{#1}\fi
\ifx\bfpage    \undefined \def\bfpage#1{#1}\fi
\ifx\blpage    \undefined \def\blpage #1{#1}\fi
\ifx\burl      \undefined \def\burl#1{\textsf{#1}}\fi
\ifx\href      \undefined \def\href#1#2{\textsf{#2}}\fi
\ifx\betal     \undefined \def\betal{\textit{et al.}}\fi
\ifx\bctitle   \undefined \def\bctitle#1{#1}\fi
\ifx\beditor   \undefined \def\beditor#1{#1}\fi
\ifx\bbtitle   \undefined \def\bbtitle#1{\textit{#1}}\fi
\ifx\bedition  \undefined \def\bedition#1{#1}\fi
\ifx\bseriesno \undefined \def\bseriesno#1{\textbf{#1}}\fi
\ifx\blocation \undefined \def\blocation#1{#1}\fi
\ifx\bsertitle \undefined \def\bsertitle#1{\textit{#1}}\fi
\ifx\bsnm      \undefined \def\bsnm#1{#1}\fi
\ifx\bsuffix   \undefined \def\bsuffix#1{#1}\fi
\ifx\bparticle \undefined \def\bparticle#1{#1}\fi
\ifx\barticle  \undefined \def\barticle#1{}\fi
\ifx\binstitute  \undefined \def\binstitute#1{#1}\fi
\ifx\bpublisher  \undefined \def\bpublisher#1{#1}\fi
\ifx\doiurl    \undefined
  \def\doiurl#1{\href{http://dx.doi.org/#1}{\textsf{DOI}}}\fi
\ifx\arxivurl  \undefined
  \def\arxivurl#1{\href{http://arxiv.org/abs/#1}{\textsf{arXiv}}}\fi
\ifx\adsurl    \undefined
  \def\adsurl#1{\href{http://adsabs.harvard.edu/abs/#1}{\textsf{ADS}}}\fi
\ifx\botherref \undefined \def\botherref#1{}\fi
\ifx\url       \undefined \def\url#1{\textsf{#1}}\fi
\ifx\bchapter  \undefined \def\bchapter#1{}\fi
\ifx\bbook     \undefined \def\bbook#1{}\fi
\ifx\bcomment  \undefined \def\bcomment#1{#1}\fi
\ifx\oauthor   \undefined \def\oauthor#1{#1}\fi
\ifx\citeauthoryear \undefined\def \citeauthoryear#1{#1}\fi
\def\endbibitem {}
\ifx\bconflocation  \undefined \def\bconflocation#1{#1} \fi

\bibitem[\protect\citeauthoryear{{Alissandrakis} and
  {Preka-Papadema}}{1984}]{ali84}
\begin{barticle}
\bauthor{\bsnm{{Alissandrakis}}, \binits{C.E.}},
\bauthor{\bsnm{{Preka-Papadema}}, \binits{P.}}:
\byear{1984},
\batitle{{Microwave emission and polarization of a flaring loop}}.
\bjtitle{\aap}
\bvolume{139},
\bfpage{507}.
\adsurl{1984A\%26A...139..507A}.
\end{barticle}
\endbibitem

\bibitem[\protect\citeauthoryear{{Altyntsev} \textit{et~al.}}{2008}]{alt08}
\begin{barticle}
\bauthor{\bsnm{{Altyntsev}}, \binits{A.T.}},
\bauthor{\bsnm{{Fleishman}}, \binits{G.D.}},
\bauthor{\bsnm{{Huang}}, \binits{G.-L.}},
\bauthor{\bsnm{{Melnikov}}, \binits{V.F.}}:
\byear{2008},
\batitle{{A broadband microwave burst produced by electron beams}}.
\bjtitle{\apj}
\bvolume{677},
\bfpage{1367}.
\doiurl{10.1086/528841}.
\adsurl{2008ApJ...677.1367A}.
\end{barticle}
\endbibitem

\bibitem[\protect\citeauthoryear{{Andries} and {Cally}}{2011}]{and11}
\begin{barticle}
\bauthor{\bsnm{{Andries}}, \binits{J.}},
\bauthor{\bsnm{{Cally}}, \binits{P.S.}}:
\byear{2011},
\batitle{{On the dispersion and scattering of magnetohydrodynamic waves by
  longitudinally stratified flux tubes}}.
\bjtitle{\apj}
\bvolume{743},
\bfpage{164}.
\doiurl{10.1088/0004-637X/743/2/164}.
\adsurl{2011ApJ...743..164A}.
\end{barticle}
\endbibitem

\bibitem[\protect\citeauthoryear{{Andries} \textit{et~al.}}{2005}]{and05}
\begin{barticle}
\bauthor{\bsnm{{Andries}}, \binits{J.}},
\bauthor{\bsnm{{Goossens}}, \binits{M.}},
\bauthor{\bsnm{{Hollweg}}, \binits{J.V.}},
\bauthor{\bsnm{{Arregui}}, \binits{I.}},
\bauthor{\bsnm{{Van Doorsselaere}}, \binits{T.}}:
\byear{2005},
\batitle{{Coronal loop oscillations. Calculation of resonantly damped MHD
  quasi-mode kink oscillations of longitudinally stratified loops}}.
\bjtitle{\aap}
\bvolume{430},
\bfpage{1109}.
\doiurl{10.1051/0004-6361:20041832}.
\adsurl{2005A\%26A...430.1109A}.
\end{barticle}
\endbibitem

\bibitem[\protect\citeauthoryear{{Antolin} and {Van
  Doorsselaere}}{2013}]{ant13}
\begin{barticle}
\bauthor{\bsnm{{Antolin}}, \binits{P.}},
\bauthor{\bsnm{{Van Doorsselaere}}, \binits{T.}}:
\byear{2013},
\batitle{{Line-of-sight geometrical and instrumental resolution effects on
  intensity perturbations by sausage modes}}.
\bjtitle{\aap}
\bvolume{555},
\bfpage{A74}.
\doiurl{10.1051/0004-6361/201220784}.
\adsurl{2013A\%26A...555A..74A}.
\end{barticle}
\endbibitem

\bibitem[\protect\citeauthoryear{{Aschwanden}}{1987}]{asc87}
\begin{barticle}
\bauthor{\bsnm{{Aschwanden}}, \binits{M.J.}}:
\byear{1987},
\batitle{{Theory of radio pulsations in coronal loops}}.
\bjtitle{\solphys}
\bvolume{111},
\bfpage{113}.
\doiurl{10.1007/BF00145445}.
\adsurl{1987SoPh..111..113A}.
\end{barticle}
\endbibitem

\bibitem[\protect\citeauthoryear{{Aschwanden} \textit{et~al.}}{1999}]{asc99}
\begin{barticle}
\bauthor{\bsnm{{Aschwanden}}, \binits{M.J.}},
\bauthor{\bsnm{{Fletcher}}, \binits{L.}},
\bauthor{\bsnm{{Schrijver}}, \binits{C.J.}},
\bauthor{\bsnm{{Alexander}}, \binits{D.}}:
\byear{1999},
\batitle{{Coronal loop oscillations observed with the Transition Region and
  Coronal Explorer}}.
\bjtitle{\apj}
\bvolume{520},
\bfpage{880}.
\doiurl{10.1086/307502}.
\adsurl{1999ApJ...520..880A}.
\end{barticle}
\endbibitem

\bibitem[\protect\citeauthoryear{{Bastian}, {Benz}, and {Gary}}{1998}]{bas98}
\begin{barticle}
\bauthor{\bsnm{{Bastian}}, \binits{T.S.}},
\bauthor{\bsnm{{Benz}}, \binits{A.O.}},
\bauthor{\bsnm{{Gary}}, \binits{D.E.}}:
\byear{1998},
\batitle{{Radio emission from solar flares}}.
\bjtitle{\araa}
\bvolume{36},
\bfpage{131}.
\doiurl{10.1146/annurev.astro.36.1.131}.
\adsurl{1998ARA\%26A..36..131B}.
\end{barticle}
\endbibitem

\bibitem[\protect\citeauthoryear{{Chen} and {Priest}}{2006}]{che06}
\begin{barticle}
\bauthor{\bsnm{{Chen}}, \binits{P.F.}},
\bauthor{\bsnm{{Priest}}, \binits{E.R.}}:
\byear{2006},
\batitle{{Transition-region explosive events: Reconnection modulated by p-mode
  waves}}.
\bjtitle{\solphys}
\bvolume{238},
\bfpage{313}.
\doiurl{10.1007/s11207-006-0215-1}.
\adsurl{2006SoPh..238..313C}.
\end{barticle}
\endbibitem

\bibitem[\protect\citeauthoryear{{Cohen}}{1960}]{coh60}
\begin{barticle}
\bauthor{\bsnm{{Cohen}}, \binits{M.H.}}:
\byear{1960},
\batitle{{Magnetoionic mode voupling at high frequencies.}}
\bjtitle{\apj}
\bvolume{131},
\bfpage{664}.
\doiurl{10.1086/146878}.
\adsurl{1960ApJ...131..664C}.
\end{barticle}
\endbibitem

\bibitem[\protect\citeauthoryear{{Costa} \textit{et~al.}}{2013}]{cos13}
\begin{barticle}
\bauthor{\bsnm{{Costa}}, \binits{J.E.R.}},
\bauthor{\bsnm{{Sim{\~o}es}}, \binits{P.J.d.A.}},
\bauthor{\bsnm{{Pinto}}, \binits{T.S.N.}},
\bauthor{\bsnm{{Melnikov}}, \binits{V.F.}}:
\byear{2013},
\batitle{{Solar burst analysis with 3D loop models}}.
\bjtitle{\pasj}
\bvolume{65},
\bfpage{5}.
\doiurl{10.1093/pasj/65.sp1.S5}.
\adsurl{2013PASJ...65S...5C}.
\end{barticle}
\endbibitem

\bibitem[\protect\citeauthoryear{{De Moortel} and {Nakariakov}}{2012}]{dem12}
\begin{barticle}
\bauthor{\bsnm{{De Moortel}}, \binits{I.}},
\bauthor{\bsnm{{Nakariakov}}, \binits{V.M.}}:
\byear{2012},
\batitle{{Magnetohydrodynamic waves and coronal seismology: an overview of
  recent results}}.
\bjtitle{Phil. Trans. Roy. Soc. London Ser. A}
\bvolume{370},
\bfpage{3193}.
\doiurl{10.1098/rsta.2011.0640}.
\adsurl{2012RSPTA.370.3193D}.
\end{barticle}
\endbibitem

\bibitem[\protect\citeauthoryear{{Doyle}, {Popescu}, and
  {Taroyan}}{2006}]{doy06}
\begin{barticle}
\bauthor{\bsnm{{Doyle}}, \binits{J.G.}},
\bauthor{\bsnm{{Popescu}}, \binits{M.D.}},
\bauthor{\bsnm{{Taroyan}}, \binits{Y.}}:
\byear{2006},
\batitle{{Repetitive occurrence of explosive events at a coronal hole
  boundary}}.
\bjtitle{\aap}
\bvolume{446},
\bfpage{327}.
\doiurl{10.1051/0004-6361:20053826}.
\adsurl{2006A\%26A...446..327D}.
\end{barticle}
\endbibitem

\bibitem[\protect\citeauthoryear{{Edwin} and {Roberts}}{1983}]{edw83}
\begin{barticle}
\bauthor{\bsnm{{Edwin}}, \binits{P.M.}},
\bauthor{\bsnm{{Roberts}}, \binits{B.}}:
\byear{1983},
\batitle{{Wave propagation in a magnetic cylinder}}.
\bjtitle{\solphys}
\bvolume{88},
\bfpage{179}.
\doiurl{10.1007/BF00196186}.
\adsurl{1983SoPh...88..179E}.
\end{barticle}
\endbibitem

\bibitem[\protect\citeauthoryear{{Erd{\'e}lyi}}{2006}]{erd06}
\begin{barticle}
\bauthor{\bsnm{{Erd{\'e}lyi}}, \binits{R.}}:
\byear{2006},
\batitle{{Magnetic coupling of waves and oscillations in the lower solar
  atmosphere: Can the tail wag the dog?}}
\bjtitle{Phil. Trans. Roy. Soc. London Ser. A}
\bvolume{364},
\bfpage{351}.
\doiurl{10.1098/rsta.2005.1703}.
\adsurl{2006RSPTA.364..351E}.
\end{barticle}
\endbibitem

\bibitem[\protect\citeauthoryear{{Fleishman} and {Kuznetsov}}{2010}]{fle10}
\begin{barticle}
\bauthor{\bsnm{{Fleishman}}, \binits{G.D.}},
\bauthor{\bsnm{{Kuznetsov}}, \binits{A.A.}}:
\byear{2010},
\batitle{{Fast gyrosynchrotron codes}}.
\bjtitle{\apj}
\bvolume{721},
\bfpage{1127}.
\doiurl{10.1088/0004-637X/721/2/1127}.
\adsurl{2010ApJ...721.1127F}.
\end{barticle}
\endbibitem

\bibitem[\protect\citeauthoryear{{Fleishman} and {Melnikov}}{2003}]{gfle03}
\begin{barticle}
\bauthor{\bsnm{{Fleishman}}, \binits{G.D.}},
\bauthor{\bsnm{{Melnikov}}, \binits{V.F.}}:
\byear{2003},
\batitle{{Gyrosynchrotron emission from anisotropic electron distributions}}.
\bjtitle{\apj}
\bvolume{587},
\bfpage{823}.
\doiurl{10.1086/368252}.
\adsurl{2003ApJ...587..823F}.
\end{barticle}
\endbibitem

\bibitem[\protect\citeauthoryear{{Fleishman}, {Bastian}, and
  {Gary}}{2008}]{gfle08}
\begin{barticle}
\bauthor{\bsnm{{Fleishman}}, \binits{G.D.}},
\bauthor{\bsnm{{Bastian}}, \binits{T.S.}},
\bauthor{\bsnm{{Gary}}, \binits{D.E.}}:
\byear{2008},
\batitle{{Broadband quasi-periodic radio and X-ray pulsations in a solar
  flare}}.
\bjtitle{\apj}
\bvolume{684},
\bfpage{1433}.
\doiurl{10.1086/589821}.
\adsurl{2008ApJ...684.1433F}.
\end{barticle}
\endbibitem

\bibitem[\protect\citeauthoryear{{Fletcher} and {Hudson}}{2008}]{fle08}
\begin{barticle}
\bauthor{\bsnm{{Fletcher}}, \binits{L.}},
\bauthor{\bsnm{{Hudson}}, \binits{H.S.}}:
\byear{2008},
\batitle{{Impulsive phase flare energy transport by large-scale Alfv{\'e}n
  waves and the electron acceleration problem}}.
\bjtitle{\apj}
\bvolume{675},
\bfpage{1645}.
\doiurl{10.1086/527044}.
\adsurl{2008ApJ...675.1645F}.
\end{barticle}
\endbibitem

\bibitem[\protect\citeauthoryear{{Gary}, {Fleishman}, and {Nita}}{2013}]{gar13}
\begin{barticle}
\bauthor{\bsnm{{Gary}}, \binits{D.E.}},
\bauthor{\bsnm{{Fleishman}}, \binits{G.D.}},
\bauthor{\bsnm{{Nita}}, \binits{G.M.}}:
\byear{2013},
\batitle{{Magnetography of solar flaring loops with microwave imaging
  spectropolarimetry}}.
\bjtitle{\solphys}
\bvolume{288},
\bfpage{549}.
\doiurl{10.1007/s11207-013-0299-3}.
\adsurl{2013SoPh..288..549G}.
\end{barticle}
\endbibitem

\bibitem[\protect\citeauthoryear{{Gruber} \textit{et~al.}}{2011}]{gru11}
\begin{barticle}
\bauthor{\bsnm{{Gruber}}, \binits{D.}},
\bauthor{\bsnm{{Lachowicz}}, \binits{P.}},
\bauthor{\bsnm{{Bissaldi}}, \binits{E.}},
\bauthor{\bsnm{{Briggs}}, \binits{M.S.}},
\bauthor{\bsnm{{Connaughton}}, \binits{V.}},
\bauthor{\bsnm{{Greiner}}, \binits{J.}},
\bauthor{\bsnm{{van der Horst}}, \binits{A.J.}},
\bauthor{\bsnm{{Kanbach}}, \binits{G.}},
\bauthor{\bsnm{{Rau}}, \binits{A.}},
\bauthor{\bsnm{{Bhat}}, \binits{P.N.}},
\bauthor{\bsnm{{Diehl}}, \binits{R.}},
\bauthor{\bsnm{{von Kienlin}}, \binits{A.}},
\bauthor{\bsnm{{Kippen}}, \binits{R.M.}},
\bauthor{\bsnm{{Meegan}}, \binits{C.A.}},
\bauthor{\bsnm{{Paciesas}}, \binits{W.S.}},
\bauthor{\bsnm{{Preece}}, \binits{R.D.}},
\bauthor{\bsnm{{Wilson-Hodge}}, \binits{C.}}:
\byear{2011},
\batitle{{Quasi-periodic pulsations in solar flares: new clues from the Fermi
  Gamma-Ray Burst Monitor}}.
\bjtitle{\aap}
\bvolume{533},
\bfpage{A61}.
\doiurl{10.1051/0004-6361/201117077}.
\adsurl{2011A\%26A...533A..61G}.
\end{barticle}
\endbibitem

\bibitem[\protect\citeauthoryear{{Holman}}{2003}]{hol03}
\begin{barticle}
\bauthor{\bsnm{{Holman}}, \binits{G.D.}}:
\byear{2003},
\batitle{{The effects of low- and high-energy cutoffs on solar flare microwave
  and hard X-ray spectra}}.
\bjtitle{\apj}
\bvolume{586},
\bfpage{606}.
\doiurl{10.1086/367554}.
\adsurl{2003ApJ...586..606H}.
\end{barticle}
\endbibitem

\bibitem[\protect\citeauthoryear{{Klein} and {Trottet}}{1984}]{kle84}
\begin{barticle}
\bauthor{\bsnm{{Klein}}, \binits{K.-L.}},
\bauthor{\bsnm{{Trottet}}, \binits{G.}}:
\byear{1984},
\batitle{{Gyrosynchrotron radiation from a source with spatially varying field
  and density}}.
\bjtitle{\aap}
\bvolume{141},
\bfpage{67}.
\adsurl{1984A\%26A...141...67K}.
\end{barticle}
\endbibitem

\bibitem[\protect\citeauthoryear{{Kliem}, {Karlick{\'y}}, and
  {Benz}}{2000}]{kli00}
\begin{barticle}
\bauthor{\bsnm{{Kliem}}, \binits{B.}},
\bauthor{\bsnm{{Karlick{\'y}}}, \binits{M.}},
\bauthor{\bsnm{{Benz}}, \binits{A.O.}}:
\byear{2000},
\batitle{{Solar flare radio pulsations as a signature of dynamic magnetic
  reconnection}}.
\bjtitle{\aap}
\bvolume{360},
\bfpage{715}.
\adsurl{2000A\%26A...360..715K}.
\end{barticle}
\endbibitem

\bibitem[\protect\citeauthoryear{{Kopylova}, {Stepanov}, and
  {Tsap}}{2002}]{kop02}
\begin{barticle}
\bauthor{\bsnm{{Kopylova}}, \binits{Y.G.}},
\bauthor{\bsnm{{Stepanov}}, \binits{A.V.}},
\bauthor{\bsnm{{Tsap}}, \binits{Y.T.}}:
\byear{2002},
\batitle{{Radial oscillations of coronal loops and microwave radiation from
  solar flares}}.
\bjtitle{Astron. Lett.}
\bvolume{28},
\bfpage{783}.
\doiurl{10.1134/1.1518717}.
\adsurl{2002AstL...28..783K}.
\end{barticle}
\endbibitem

\bibitem[\protect\citeauthoryear{{Kucera} \textit{et~al.}}{1993}]{kuc93}
\begin{barticle}
\bauthor{\bsnm{{Kucera}}, \binits{T.A.}},
\bauthor{\bsnm{{Dulk}}, \binits{G.A.}},
\bauthor{\bsnm{{Kiplinger}}, \binits{A.L.}},
\bauthor{\bsnm{{Winglee}}, \binits{R.M.}},
\bauthor{\bsnm{{Bastian}}, \binits{T.S.}},
\bauthor{\bsnm{{Graeter}}, \binits{M.}}:
\byear{1993},
\batitle{{Multiple wavelength observations of an off-limb eruptive solar
  flare}}.
\bjtitle{\apj}
\bvolume{412},
\bfpage{853}.
\doiurl{10.1086/172967}.
\adsurl{1993ApJ...412..853K}.
\end{barticle}
\endbibitem

\bibitem[\protect\citeauthoryear{{Kundu} \textit{et~al.}}{2001}]{kun01}
\begin{barticle}
\bauthor{\bsnm{{Kundu}}, \binits{M.R.}},
\bauthor{\bsnm{{Nindos}}, \binits{A.}},
\bauthor{\bsnm{{White}}, \binits{S.M.}},
\bauthor{\bsnm{{Grechnev}}, \binits{V.V.}}:
\byear{2001},
\batitle{{A multiwavelength study of three solar flares}}.
\bjtitle{\apj}
\bvolume{557},
\bfpage{880}.
\doiurl{10.1086/321534}.
\adsurl{2001ApJ...557..880K}.
\end{barticle}
\endbibitem

\bibitem[\protect\citeauthoryear{{Kupriyanova} \textit{et~al.}}{2010}]{kup10}
\begin{barticle}
\bauthor{\bsnm{{Kupriyanova}}, \binits{E.G.}},
\bauthor{\bsnm{{Melnikov}}, \binits{V.F.}},
\bauthor{\bsnm{{Nakariakov}}, \binits{V.M.}},
\bauthor{\bsnm{{Shibasaki}}, \binits{K.}}:
\byear{2010},
\batitle{{Types of microwave quasi-periodic pulsations in single flaring
  loops}}.
\bjtitle{\solphys}
\bvolume{267},
\bfpage{329}.
\doiurl{10.1007/s11207-010-9642-0}.
\adsurl{2010SoPh..267..329K}.
\end{barticle}
\endbibitem

\bibitem[\protect\citeauthoryear{{Kuznetsov} and {Kontar}}{2015}]{kuz14}
\begin{barticle}
\bauthor{\bsnm{{Kuznetsov}}, \binits{A.A.}},
\bauthor{\bsnm{{Kontar}}, \binits{E.P.}}:
\byear{2015},
\batitle{{Spatially resolved energetic electron properties for the 21 May 2004
  flare from radio observations and 3D simulations}}.
\bjtitle{\solphys}
\bvolume{290},
\bfpage{79}.
\doiurl{10.1007/s11207-014-0530-x}.
\adsurl{2015SoPh..290...79K}.
\end{barticle}
\endbibitem

\bibitem[\protect\citeauthoryear{{Kuznetsov} and {Zharkova}}{2010}]{kuz10}
\begin{barticle}
\bauthor{\bsnm{{Kuznetsov}}, \binits{A.A.}},
\bauthor{\bsnm{{Zharkova}}, \binits{V.V.}}:
\byear{2010},
\batitle{{Manifestations of energetic electrons with anisotropic distributions
  in solar flares. II. Gyrosynchrotron microwave emission}}.
\bjtitle{\apj}
\bvolume{722},
\bfpage{1577}.
\doiurl{10.1088/0004-637X/722/2/1577}.
\adsurl{2010ApJ...722.1577K}.
\end{barticle}
\endbibitem

\bibitem[\protect\citeauthoryear{{Kuznetsov}, {Nita}, and
  {Fleishman}}{2011}]{kuz11}
\begin{barticle}
\bauthor{\bsnm{{Kuznetsov}}, \binits{A.A.}},
\bauthor{\bsnm{{Nita}}, \binits{G.M.}},
\bauthor{\bsnm{{Fleishman}}, \binits{G.D.}}:
\byear{2011},
\batitle{{Three-dimensional simulations of gyrosynchrotron emission from mildly
  anisotropic nonuniform electron distributions in symmetric magnetic loops}}.
\bjtitle{\apj}
\bvolume{742},
\bfpage{87}.
\doiurl{10.1088/0004-637X/742/2/87}.
\adsurl{http://esoads.eso.org/abs/2011ApJ...742...87K}.
\end{barticle}
\endbibitem

\bibitem[\protect\citeauthoryear{{Kuznetsov} and {Melnikov}}{2012}]{skuz12}
\begin{barticle}
\bauthor{\bsnm{{Kuznetsov}}, \binits{S.A.}},
\bauthor{\bsnm{{Melnikov}}, \binits{V.F.}}:
\byear{2012},
\batitle{{Modeling the effect of plasma density on the dynamics of the
  microwave spectrum of solar flaring loops}}.
\bjtitle{Geomagn. Aer.}
\bvolume{52},
\bfpage{883}.
\doiurl{10.1134/S0016793212070092}.
\adsurl{2012Ge\%26Ae..52..883K}.
\end{barticle}
\endbibitem

\bibitem[\protect\citeauthoryear{{Melnikov} \textit{et~al.}}{2005}]{mel05}
\begin{barticle}
\bauthor{\bsnm{{Melnikov}}, \binits{V.F.}},
\bauthor{\bsnm{{Reznikova}}, \binits{V.E.}},
\bauthor{\bsnm{{Shibasaki}}, \binits{K.}},
\bauthor{\bsnm{{Nakariakov}}, \binits{V.M.}}:
\byear{2005},
\batitle{{Spatially resolved microwave pulsations of a flare loop}}.
\bjtitle{\aap}
\bvolume{439},
\bfpage{727}.
\doiurl{10.1051/0004-6361:20052774}.
\adsurl{2005A\%26A...439..727M}.
\end{barticle}
\endbibitem

\bibitem[\protect\citeauthoryear{{Melrose}}{1968}]{mel68}
\begin{barticle}
\bauthor{\bsnm{{Melrose}}, \binits{D.B.}}:
\byear{1968},
\batitle{{The emission and absorption of waves by charged particles in
  magnetized plasmas}}.
\bjtitle{\apss}
\bvolume{2},
\bfpage{171}.
\doiurl{10.1007/BF00651567}.
\adsurl{1968Ap\%26SS...2..171M}.
\end{barticle}
\endbibitem

\bibitem[\protect\citeauthoryear{{Mossessian} and {Fleishman}}{2012}]{mos12}
\begin{barticle}
\bauthor{\bsnm{{Mossessian}}, \binits{G.}},
\bauthor{\bsnm{{Fleishman}}, \binits{G.D.}}:
\byear{2012},
\batitle{{Modeling of gyrosynchrotron radio emission pulsations produced by
  magnetohydrodynamic loop oscillations in solar flares}}.
\bjtitle{\apj}
\bvolume{748},
\bfpage{140}.
\doiurl{10.1088/0004-637X/748/2/140}.
\adsurl{2012ApJ...748..140M}.
\end{barticle}
\endbibitem

\bibitem[\protect\citeauthoryear{{Nakariakov} and {Melnikov}}{2006}]{nak06b}
\begin{barticle}
\bauthor{\bsnm{{Nakariakov}}, \binits{V.M.}},
\bauthor{\bsnm{{Melnikov}}, \binits{V.F.}}:
\byear{2006},
\batitle{{Modulation of gyrosynchrotron emission in solar and stellar flares by
  slow magnetoacoustic oscillations}}.
\bjtitle{\aap}
\bvolume{446},
\bfpage{1151}.
\doiurl{10.1051/0004-6361:20053944}.
\adsurl{2006A\%26A...446.1151N}.
\end{barticle}
\endbibitem

\bibitem[\protect\citeauthoryear{{Nakariakov} and {Melnikov}}{2009}]{nak09}
\begin{barticle}
\bauthor{\bsnm{{Nakariakov}}, \binits{V.M.}},
\bauthor{\bsnm{{Melnikov}}, \binits{V.F.}}:
\byear{2009},
\batitle{{Quasi-periodic pulsations in solar flares}}.
\bjtitle{\ssr}
\bvolume{149},
\bfpage{119}.
\doiurl{10.1007/s11214-009-9536-3}.
\adsurl{2009SSRv..149..119N}.
\end{barticle}
\endbibitem

\bibitem[\protect\citeauthoryear{{Nakariakov} and {Verwichte}}{2005}]{nak05}
\begin{barticle}
\bauthor{\bsnm{{Nakariakov}}, \binits{V.M.}},
\bauthor{\bsnm{{Verwichte}}, \binits{E.}}:
\byear{2005},
\batitle{{Coronal waves and oscillations}}.
\bjtitle{Living Rev. Solar Phys.}
\bvolume{2},
\bfpage{3}.
\doiurl{10.12942/lrsp-2005-3}.
\adsurl{2005LRSP....2....3N}.
\burl{http://solarphysics.livingreviews.org/Articles/lrsp-2005-3/}.
\end{barticle}
\endbibitem

\bibitem[\protect\citeauthoryear{{Nakariakov}, {Kashapova}, and
  {Yan}}{2014}]{nak14}
\begin{barticle}
\bauthor{\bsnm{{Nakariakov}}, \binits{V.M.}},
\bauthor{\bsnm{{Kashapova}}, \binits{L.K.}},
\bauthor{\bsnm{{Yan}}, \binits{Y.-H.}}:
\byear{2014},
\batitle{{Editorial: solar radiophysics -- recent results on observations and
  theories}}.
\bjtitle{Res. Astron. Astrophys.}
\bvolume{14},
\bfpage{1}.
\doiurl{10.1088/1674-4527/14/7/001}.
\adsurl{2014RAA....14....1N}.
\end{barticle}
\endbibitem

\bibitem[\protect\citeauthoryear{{Nakariakov}, {Melnikov}, and
  {Reznikova}}{2003}]{nak03}
\begin{barticle}
\bauthor{\bsnm{{Nakariakov}}, \binits{V.M.}},
\bauthor{\bsnm{{Melnikov}}, \binits{V.F.}},
\bauthor{\bsnm{{Reznikova}}, \binits{V.E.}}:
\byear{2003},
\batitle{{Global sausage modes of coronal loops}}.
\bjtitle{\aap}
\bvolume{412},
\bfpage{L7}.
\doiurl{10.1051/0004-6361:20031660}.
\adsurl{2003A\%26A...412L...7N}.
\end{barticle}
\endbibitem

\bibitem[\protect\citeauthoryear{{Nakariakov} \textit{et~al.}}{1999}]{nak99}
\begin{barticle}
\bauthor{\bsnm{{Nakariakov}}, \binits{V.M.}},
\bauthor{\bsnm{{Ofman}}, \binits{L.}},
\bauthor{\bsnm{{Deluca}}, \binits{E.E.}},
\bauthor{\bsnm{{Roberts}}, \binits{B.}},
\bauthor{\bsnm{{Davila}}, \binits{J.M.}}:
\byear{1999},
\batitle{{TRACE observation of damped coronal loop oscillations: Implications
  for coronal heating}}.
\bjtitle{Science}
\bvolume{285},
\bfpage{862}.
\doiurl{10.1126/science.285.5429.862}.
\adsurl{1999Sci...285..862N}.
\end{barticle}
\endbibitem

\bibitem[\protect\citeauthoryear{{Nakariakov} \textit{et~al.}}{2006}]{nak06}
\begin{barticle}
\bauthor{\bsnm{{Nakariakov}}, \binits{V.M.}},
\bauthor{\bsnm{{Foullon}}, \binits{C.}},
\bauthor{\bsnm{{Verwichte}}, \binits{E.}},
\bauthor{\bsnm{{Young}}, \binits{N.P.}}:
\byear{2006},
\batitle{{Quasi-periodic modulation of solar and stellar flaring emission by
  magnetohydrodynamic oscillations in a nearby loop}}.
\bjtitle{\aap}
\bvolume{452},
\bfpage{343}.
\doiurl{10.1051/0004-6361:20054608}.
\adsurl{2006A\%26A...452..343N}.
\end{barticle}
\endbibitem

\bibitem[\protect\citeauthoryear{{Nindos} \textit{et~al.}}{2000}]{nin00}
\begin{barticle}
\bauthor{\bsnm{{Nindos}}, \binits{A.}},
\bauthor{\bsnm{{White}}, \binits{S.M.}},
\bauthor{\bsnm{{Kundu}}, \binits{M.R.}},
\bauthor{\bsnm{{Gary}}, \binits{D.E.}}:
\byear{2000},
\batitle{{Observations and models of a flaring loop}}.
\bjtitle{\apj}
\bvolume{533},
\bfpage{1053}.
\doiurl{10.1086/308705}.
\adsurl{2000ApJ...533.1053N}.
\end{barticle}
\endbibitem

\bibitem[\protect\citeauthoryear{{Nita} \textit{et~al.}}{2015}]{nit14}
\begin{barticle}
\bauthor{\bsnm{{Nita}}, \binits{G.M.}},
\bauthor{\bsnm{{Fleishman}}, \binits{G.D.}},
\bauthor{\bsnm{{Kuznetsov}}, \binits{A.A.}},
\bauthor{\bsnm{{Kontar}}, \binits{E.P.}},
\bauthor{\bsnm{{Gary}}, \binits{D.E.}}:
\byear{2015},
\batitle{{Three-dimensional radio and X-ray modeling and data analysis
  software: Revealing flare complexity}}.
\bjtitle{\apj}
\bvolume{799},
\bfpage{236}.
\doiurl{10.1088/0004-637X/799/2/236}.
\adsurl{2015ApJ...799..236N}.
\end{barticle}
\endbibitem

\bibitem[\protect\citeauthoryear{{Preka-Papadema} and
  {Alissandrakis}}{1988}]{pre88}
\begin{barticle}
\bauthor{\bsnm{{Preka-Papadema}}, \binits{P.}},
\bauthor{\bsnm{{Alissandrakis}}, \binits{C.E.}}:
\byear{1988},
\batitle{{Spatial and spectral structure of a solar flaring loop at centimeter
  wavelengths}}.
\bjtitle{\aap}
\bvolume{191},
\bfpage{365}.
\adsurl{1988A\%26A...191..365P}.
\end{barticle}
\endbibitem

\bibitem[\protect\citeauthoryear{{Preka-Papadema} and
  {Alissandrakis}}{1992}]{pre92}
\begin{barticle}
\bauthor{\bsnm{{Preka-Papadema}}, \binits{P.}},
\bauthor{\bsnm{{Alissandrakis}}, \binits{C.E.}}:
\byear{1992},
\batitle{{Two-dimensional model maps of flaring loops at cm-wavelengths}}.
\bjtitle{\aap}
\bvolume{257},
\bfpage{307}.
\adsurl{1992A\%26A...257..307P}.
\end{barticle}
\endbibitem

\bibitem[\protect\citeauthoryear{{Ramaty}}{1969}]{ram69}
\begin{barticle}
\bauthor{\bsnm{{Ramaty}}, \binits{R.}}:
\byear{1969},
\batitle{{Gyrosynchrotron emission and absorption in a magnetoactive plasma}}.
\bjtitle{\apj}
\bvolume{158},
\bfpage{753}.
\doiurl{10.1086/150235}.
\adsurl{1969ApJ...158..753R}.
\end{barticle}
\endbibitem

\bibitem[\protect\citeauthoryear{{Razin}}{1960a}]{raz60b}
\begin{barticle}
\bauthor{\bsnm{{Razin}}, \binits{V.A.}}:
\byear{1960}a,
\batitle{{On the spectrum of nonthermal cosmic radio emission}}.
\bjtitle{Izv. Vyssh. Uchebn. Zaved. Radiofiz.}
\bvolume{3},
\bfpage{921}.
\end{barticle}
\endbibitem

\bibitem[\protect\citeauthoryear{{Razin}}{1960b}]{raz60a}
\begin{barticle}
\bauthor{\bsnm{{Razin}}, \binits{V.A.}}:
\byear{1960}b,
\batitle{{On the theory of radio spectra of discrete sources at the frequencies
  below 30 MHz}}.
\bjtitle{Izv. Vyssh. Uchebn. Zaved. Radiofiz.}
\bvolume{3},
\bfpage{584}.
\end{barticle}
\endbibitem

\bibitem[\protect\citeauthoryear{{Reznikova}, {Antolin}, and {Van
  Doorsselaere}}{2014}]{rez14}
\begin{barticle}
\bauthor{\bsnm{{Reznikova}}, \binits{V.E.}},
\bauthor{\bsnm{{Antolin}}, \binits{P.}},
\bauthor{\bsnm{{Van Doorsselaere}}, \binits{T.}}:
\byear{2014},
\batitle{{Forward modeling of gyrosynchrotron intensity perturbations by
  sausage modes}}.
\bjtitle{\apj}
\bvolume{785},
\bfpage{86}.
\doiurl{10.1088/0004-637X/785/2/86}.
\adsurl{2014ApJ...785...86R}.
\end{barticle}
\endbibitem

\bibitem[\protect\citeauthoryear{{Reznikova}, {Van Doorsselaere}, and
  {Kuznetsov}}{2015}]{rez15}
\begin{botherref}
\oauthor{\bsnm{{Reznikova}}, \binits{V.E.}},
\oauthor{\bsnm{{Van Doorsselaere}}, \binits{T.}},
\oauthor{\bsnm{{Kuznetsov}}, \binits{A.A.}}:
2015,
{Perturbations of gyrosynchrotron emission polarization from solar flares by
  sausage modes: Forward modeling}.
\textit{\aap, {\rm in press}}.
\doiurl{10.1051/0004-6361/201424548}.
\end{botherref}
\endbibitem

\bibitem[\protect\citeauthoryear{{Reznikova} \textit{et~al.}}{2007}]{rez07}
\begin{barticle}
\bauthor{\bsnm{{Reznikova}}, \binits{V.E.}},
\bauthor{\bsnm{{Melnikov}}, \binits{V.F.}},
\bauthor{\bsnm{{Su}}, \binits{Y.}},
\bauthor{\bsnm{{Huang}}, \binits{G.}}:
\byear{2007},
\batitle{{Pulsations of microwave flaring emission at low and high
  frequencies}}.
\bjtitle{Astronomy Reports}
\bvolume{51},
\bfpage{588}.
\doiurl{10.1134/S1063772907070086}.
\adsurl{2007ARep...51..588R}.
\end{barticle}
\endbibitem

\bibitem[\protect\citeauthoryear{{Reznikova} \textit{et~al.}}{2009}]{rez09}
\begin{barticle}
\bauthor{\bsnm{{Reznikova}}, \binits{V.E.}},
\bauthor{\bsnm{{Melnikov}}, \binits{V.F.}},
\bauthor{\bsnm{{Shibasaki}}, \binits{K.}},
\bauthor{\bsnm{{Gorbikov}}, \binits{S.P.}},
\bauthor{\bsnm{{Pyatakov}}, \binits{N.P.}},
\bauthor{\bsnm{{Myagkova}}, \binits{I.N.}},
\bauthor{\bsnm{{Ji}}, \binits{H.}}:
\byear{2009},
\batitle{{2002 August 24 limb flare loop: Dynamics of microwave brightness
  distribution}}.
\bjtitle{\apj}
\bvolume{697},
\bfpage{735}.
\doiurl{10.1088/0004-637X/697/1/735}.
\adsurl{2009ApJ...697..735R}.
\end{barticle}
\endbibitem

\bibitem[\protect\citeauthoryear{{Sim{\~o}es} and {Costa}}{2006}]{sim06}
\begin{barticle}
\bauthor{\bsnm{{Sim{\~o}es}}, \binits{P.J.A.}},
\bauthor{\bsnm{{Costa}}, \binits{J.E.R.}}:
\byear{2006},
\batitle{{Solar bursts gyrosynchrotron emission from three-dimensional
  sources}}.
\bjtitle{\aap}
\bvolume{453},
\bfpage{729}.
\doiurl{10.1051/0004-6361:20054665}.
\adsurl{2006A\%26A...453..729S}.
\end{barticle}
\endbibitem

\bibitem[\protect\citeauthoryear{{Sim{\~o}es} and {Costa}}{2010}]{sim10}
\begin{barticle}
\bauthor{\bsnm{{Sim{\~o}es}}, \binits{P.J.A.}},
\bauthor{\bsnm{{Costa}}, \binits{J.E.R.}}:
\byear{2010},
\batitle{{Gyrosynchrotron emission from anisotropic pitch-angle distribution of
  electrons in 3-D solar flare sources}}.
\bjtitle{\solphys}
\bvolume{266},
\bfpage{109}.
\doiurl{10.1007/s11207-010-9596-2}.
\adsurl{2010SoPh..266..109S}.
\end{barticle}
\endbibitem

\bibitem[\protect\citeauthoryear{{Tzatzakis}, {Nindos}, and
  {Alissandrakis}}{2008}]{tza08}
\begin{barticle}
\bauthor{\bsnm{{Tzatzakis}}, \binits{V.}},
\bauthor{\bsnm{{Nindos}}, \binits{A.}},
\bauthor{\bsnm{{Alissandrakis}}, \binits{C.E.}}:
\byear{2008},
\batitle{{A statistical study of microwave flare morphologies}}.
\bjtitle{\solphys}
\bvolume{253},
\bfpage{79}.
\doiurl{10.1007/s11207-008-9263-z}.
\adsurl{2008SoPh..253...79T}.
\end{barticle}
\endbibitem

\bibitem[\protect\citeauthoryear{{Van Doorsselaere}
  \textit{et~al.}}{2004}]{vand04}
\begin{barticle}
\bauthor{\bsnm{{Van Doorsselaere}}, \binits{T.}},
\bauthor{\bsnm{{Debosscher}}, \binits{A.}},
\bauthor{\bsnm{{Andries}}, \binits{J.}},
\bauthor{\bsnm{{Poedts}}, \binits{S.}}:
\byear{2004},
\batitle{{The effect of curvature on quasi-modes in coronal loops}}.
\bjtitle{\aap}
\bvolume{424},
\bfpage{1065}.
\doiurl{10.1051/0004-6361:20041239}.
\adsurl{2004A\%26A...424.1065V}.
\end{barticle}
\endbibitem

\bibitem[\protect\citeauthoryear{{Zaitsev} and {Stepanov}}{1975}]{zaj75}
\begin{barticle}
\bauthor{\bsnm{{Zaitsev}}, \binits{V.V.}},
\bauthor{\bsnm{{Stepanov}}, \binits{A.V.}}:
\byear{1975},
\batitle{{On the origin of pulsations of type IV solar radio emission. Plasma
  cylinder oscillations (I).}}
\bjtitle{Issled. Geomagn. Aeron. Fiz. Solntsa}
\bvolume{37},
\bfpage{3}.
\adsurl{1975IGAFS..37....3Z}.
\end{barticle}
\endbibitem

\bibitem[\protect\citeauthoryear{{Zheleznyakov} and {Zlotnik}}{1964}]{zhe64}
\begin{barticle}
\bauthor{\bsnm{{Zheleznyakov}}, \binits{V.V.}},
\bauthor{\bsnm{{Zlotnik}}, \binits{E.Y.}}:
\byear{1964},
\batitle{{Polarization of radio waves passing through a transverse magnetic
  field region in the solar corona}}.
\bjtitle{\sovast}
\bvolume{7},
\bfpage{485}.
\adsurl{1964SvA.....7..485Z}.
\end{barticle}
\endbibitem

\end{thebibliography}
\end{document}